\documentclass[11pt]{article} 
\usepackage[top=20mm,bottom=20mm, left=20mm, right=20mm]{geometry}
\usepackage{authblk,graphicx,amsmath,amssymb,url,enumerate,mathrsfs,epsfig,color,transparent,cite}
\usepackage{hyperref,caption,subcaption}
 \hypersetup{
     colorlinks=true,       
     linkcolor=red,          
     citecolor=blue,        
     filecolor=magenta,      
     urlcolor=cyan           
 }

 \numberwithin{equation}{section}
 
\newcommand{\op}[1]{\skew{5}\hat{#1}}

\title{Defects and Metric Anomalies in F{\"o}ppl-von K{\'a}rm{\'a}n Surfaces}
\author[1]{Manish Singh}
\author[2]{Ayan Roychowdhury}
\author[1]{Anurag Gupta\thanks{ag@iitk.ac.in}}
\affil[1]{Department of Mechanical Engineering, Indian Institute of Technology Kanpur, 208016, India}
\affil[2]{Simons Centre for the Study of Living Machines, National Centre for Biological Sciences, Bangalore, 560065, India}

\date{\today}

\begin{document}
\maketitle

\begin{abstract}
A general framework is developed to study the deformation and stress response in F{\"o}ppl-von K{\'a}rm{\'a}n shallow shells for a given distribution of defects, such as dislocations, disclinations, and interstitials, and metric anomalies, such as thermal and growth strains. The theory includes dislocations and disclinations whose defect lines can both pierce the two-dimensional surface and lie within the surface. An essential aspect of the theory is the derivation of strain incompatibility relations for stretching and bending strains with incompatibility sources in terms of various defect and metric anomaly densities. The incompatibility relations are combined with balance laws and constitutive assumptions to obtain the inhomogeneous F{\"o}ppl-von K{\'a}rm{\'a}n equations for shallow shells.  Several boundary value problems are posed, and solved numerically, by first considering only dislocations and then disclinations coupled with growth strains.
\end{abstract}


\section{Introduction}
Defects are local disruptions of translational and rotational material order while metric anomalies (e.g., due to thermal, growth, and hygroelastic strains) are local disruptions of the metrical order. Both play an important role in influencing geometrical (shape, topology, etc.) and physical (mechanical, electrical, etc.) properties of two-dimensional (2D) elastic surfaces (e.g., 2D materials and biological membranes)~\cite{harris70, harris74, nelson-book, bowickgiomi09, liang-mahadevan11, rcgupta20}. In the context of elastic sheets, defects and metric anomalies act as internal sources for deformation and stress in the 2D surface. The problem of interest therefore is to evaluate the deformed shape and the stress distribution, associated with the elastic surface, for a given prescription of defects and metric anomalies. The problem has classically attracted attention for defects in thin plates within the framework of small deformation linearized elasticity~\cite{eshelbystroh51, mitchell61, saitoetal72, nabarrokostlan78, eshelby79}. Although analytical solutions were variously obtained, they were limited in their scope due to the assumption of small deformation. The problem was subsequently dealt with for defects in F{\"o}ppl-von K{\'a}rm{\'a}n plates allowing for moderate bending but small stretching of the surface~\cite{SeungNelson88, nelson-book}. The resulting solutions have been useful in understanding a wide range of phenomena in solid condensed matter literature~\cite{nelson-book, bowickgiomi09} and in defect controlled topological designing of 2D materials~\cite{Gao14}.  The theory was however restricted to  isotropic point defects and isolated wedge disclinations and edge dislocations whose defect lines pierce normally through the surface. The other types of line defects (with arbitrary line directions and Burgers/Frank vector), as well as defects in the form of distributed densities over the surface and anisotropic point defects, have not been discussed and cannot be incorporated in the existing framework in any straightforward manner unless the geometrical nature of defects is developed systematically (as has been pursued in the present work). 
The problem of metric anomalies in F{\"o}ppl-von K{\'a}rm{\'a}n plates (and shells) is better understood and has been used, for instance, to study the emergent shape of certain botanical objects~\cite{liang-mahadevan11}. A comprehensive theory which combines defects and metric anomalies is however lacking altogether (except for our recent attempt~\cite{rcgupta17}). Such a theory holds promise for modelling surfaces where defects coexist with thermal/growth strains and can either enhance or screen each other's influence on the mechanics of the surface. 

The present work is concerned with formulating a general theory for modelling the mechanical response of 2D elastic surfaces, considered as F{\"o}ppl-von K{\'a}rm{\'a}n shallow shells, in the presence of dislocations, disclinations, point defects, and metric anomalies. In doing so, we fill the aforementioned gaps and generalize existing frameworks. In particular, we allow for dislocations and disclinations whose defect lines are oriented arbitrarily with respect to the surface, and also for metric anomalies to appear coupled with the defect densities. Our methodology is based on a geometric theory of defects and the notion of a geometric configuration (of the surface) which is not embeddable in $\mathbb{R}^3$ due to its inherent defectiveness. In other words, the defective configuration fails to satisfy the Gauss and Codazzi-Mainardi compatibility equations. This loss of compatibility is closely related to the emergence of incompatible strain fields throughout the surface. Similar ideas have been used successfully to develop micromechanical theories of defects in small deformation linear elastic solids~\cite{kroner81a, rcgupta16} and three-dimensional (3D) nonlinear elastic solids~\cite{arash}. Although our geometric description of defects is general (within Kirchhoff Love kinematics), we pose our boundary value problems within a more restrictive F{\"o}ppl-von K{\'a}rm{\'a}n framework which, while incorporating geometrical nonlinearities, allows us to decompose total stretching and bending strains additively into elastic and anelastic parts in addition to working with a scalar stress function, rather than a stress tensor, and a scalar transverse deformation field, rather than a displacement vector.

We note that there have been significant developments in the recent literature on incompatible plates/shells within a more general kinematical framework than assumed in our work~\cite{efrati09, sadik16}; these works are however restricted to incorporating incompatibility arising only due to growth in the 2D domain. Their applicability in developing complete boundary value problems to study the micromechanics of defects remains open. Similarly, there is significant theoretical and numerical work in the context of non-linear elasto-plastic deformations (including shells)~\cite{srini98, simo, simobook}. Our interest however is in solving for deformation and stress fields given a distribution of defects (rather than incompatibility or plastic strains). The defect distribution is both physically observable and experimentally quantifiable (clearly so for 2D materials~\cite{bowickgiomi09}) and it is of much interest to pose boundary value problems with defects appearing explicitly as sources of deformation and stress~\cite{kroner81a}.

There are three central contributions of this paper. Among these are two theoretical results. The first result is a system of strain incompatibility relations, given in Equations \eqref{s-i-r-fvk}, which explicitly demonstrates the loss of compatibility in anelastic stretching and bending strains due to a distribution of defects and metric anomalies. The second result is the pair of inhomogeneous F{\"o}ppl-von K{\'a}rm{\'a}n relations with dislocation and disclination densities as given in Equations \eqref{Firstvk-onlyd}-\eqref{Secvk-onlyd} and \eqref{Firstvk-discl}-\eqref{Secvk-discl}, respectively, the latter inclusive of a coupling with growth strains. These are the required governing equations for the determination of transverse deformation and stress fields associated with the elastic surface due to the presence of respective defect distributions. The third contribution is the numerical solution to several boundary value problems involving various configurations of dislocations and wedge disclinations with isotropic growth strains in F{\"o}ppl-von K{\'a}rm{\'a}n plates (see Sections~\ref{dislocations} and \ref{disclinations}). The former include isolated edge dislocations with their defect line and Burgers vector both lying within the plate surface. These examples provide novel demonstrations of folding an elastic sheet into a variety of configurations. The latter set of problems study the coupling of an isolated disclination with multiple combinations of surface and bending growth strains. In doing so we emphasize the morphological richness which emerges from the coupling between disclinations and growth. Our framework is immediately applicable for understanding the mechanical properties of defects in thin films~\cite{Pandey21}, for providing novel avenues in the paper folding problems, and for developing insights into the problem of growth of biological membranes assisted by defects.

\section{Kinematics and geometry of defects}
\label{kinematics}

Let $\omega$ be a 2D simply connected bounded manifold with piecewise smooth boundary, homeomorphic to the closed  disc in $\mathbb{R}^2$, and let $\mathscr{B}=\omega\times[-h/2,h/2]$, for some real constant $h>0$, be the cylindrical closed neighbourhood of $\omega$. We assume $\mathscr{B}$ to represent a sufficiently thin 3D shell with a mid-surface as $\omega$. The thinness of the shell is quantified using $\epsilon=({h}/{L})\ll 1$, where $L$ is a characteristic linear dimension of the mid-surface~\cite{koiter1966}.  The shell kinematics is assumed to satisfy the Kirchhoff-Love hypothesis which requires all the planar sections orthogonal to the mid-surface $\omega$ of the shell $\mathscr{B}$  to remain planar and orthogonal to the mid-surface  under all sufficiently smooth deformations of $\mathscr{B}$; tilting of the transverse directions and thickness distentions are therefore not permitted.
Let $(\theta^1, \theta^2)$ be the natural coordinate system on $\omega$, and let $\theta^3 = \zeta$ be the transverse coordinate along the thickness direction. The adapted coordinates $(\theta^\alpha,\zeta)$ are assumed to be convected by all the embeddings of $\omega$. The small case Greek indices $\alpha,\,\beta,\,\mu\,\ldots$ etc., and  the small case Roman indices $i,\,j,\,k\,\ldots$ etc., take values from the sets $\{1,2\}$ and $\{1,2,3\}$, respectively. A subscript comma is used to denote ordinary spatial derivatives with respect to the natural coordinates $\theta^i$. 

The inner product, cross product, and tensor product, in 3D Euclidean vector spaces, are denoted by $\cdot$, $\times$, and $\otimes$, respectively. The 2D and 3D permutational symbols are represented as $e_{\alpha\beta}$ and $e_{ijk}$, respectively. The Kronecker delta symbols $\delta_\alpha^\beta$, $\delta_{\alpha \beta}$, and $\delta^{\alpha \beta}$ carry the usual meaning. Round and square brackets enclosing indices in the subscript are used to denote, respectively, symmetrization and anti-symmetrization with respect to the enclosed indices, i.e.,  $C_{(ij)}=(C_{ij}+C_{ji})/2$, $C_{[ij]}=(C_{ij}-C_{ji})/2$, etc. The enclosed indices within two vertical bars in the subscript are to be exempted from anti-symmetrization; e.g., $2A_{[i|jk|l]} = A_{ijkl} - A_{ljki}$. We say that a function $f$ is of order $O(\epsilon^r)$ if and only if there exist positive constants $M$ and $\delta$ such that $| f | \leq  M |\epsilon|^r$ for all $\epsilon < \delta$, where $r$ is any real number. A function $f$ is of order $o(\epsilon^r)$ if $f/\epsilon^r \rightarrow 0$ as $\epsilon \rightarrow 0$. We use $[A_{ij}]$ to represent the matrix with components $A_{ij}$ and $\text{det}[A_{ij}]$ as the associated determinant. Let $f(\theta^\alpha)$ and $g(\theta^\alpha)$ be two fields defined over the surface. The 2D Laplacian and biharmonic operators $\Delta$ and  $\Delta^2$ are defined such that $\Delta f = f_{,11} + f_{,22}$ and $\Delta^2 f=f_{,1111}+2f_{,1122}+f_{,2222}$, respectively.  The 2D Monge-Amp{\`e}re bracket $[\cdot,\cdot]$ is defined by $[f,g]=e^{\alpha\beta}e^{\mu\nu}f_{,\alpha\mu}g_{,\beta\nu}=f_{,11}g_{,22} +f_{,22}g_{,11} -2f_{,12}g_{,12}$; in particular $[f,f]=2f_{,11}f_{,22} -2f_{,12}f_{,12} = 2 \text{det} [f_{,\alpha \beta}]$. Here, and elsewhere in the paper, the fields are assumed to have the required number of continuous derivatives.

\subsection{The reference and the current configuration}
Let a fixed reference configuration of the shell mid-surface be given by an isometric embedding $\boldsymbol{R}:\omega\to\mathbb{R}^3$ whose tangent spaces are spanned by the natural basis vectors $\boldsymbol{A}_\alpha=\boldsymbol{R}_{,\alpha}$. The first and second fundamental forms associated with the reference surface $\boldsymbol{R}(\omega)$ are $A_{\alpha\beta}=\boldsymbol{A}_\alpha{\cdot}\boldsymbol{A}_\beta$ and $B_{\alpha\beta}=-\boldsymbol{N}_{,\alpha}{\cdot}\boldsymbol{A}_\beta$, respectively, where $\boldsymbol{N}={\boldsymbol{A}_1\times\boldsymbol{A}_2}/{|\boldsymbol{A}_1\times\boldsymbol{A}_2|}$ is the unit normal field.  The current (deformed) configuration of the shell mid-surface is given by an isometric embedding $\op{\boldsymbol{R}}:\omega\to\mathbb{R}^3$ with tangent spaces spanned by the natural basis vectors $\op{\boldsymbol{A}}_\alpha=\op{\boldsymbol{R}}_{,\alpha}$. The associated  first and second fundamental forms are $\hat{A}_{\alpha\beta}=\op{\boldsymbol{A}}_\alpha{\cdot}\op{\boldsymbol{A}}_\beta$ and $\hat{B}_{\alpha\beta}=-\op{\boldsymbol{N}}_{,\alpha}{\cdot}\op{\boldsymbol{A}}_\beta$, respectively, where $\hat{\boldsymbol{N}}={\op{\boldsymbol{A}}_1\times\op{\boldsymbol{A}}_2}/{|\op{\boldsymbol{A}}_1\times\op{\boldsymbol{A}}_2|}$ is the unit normal field on the deformed surface. The pairs $(A_{\alpha\beta},B_{\alpha\beta})$ and $(\hat{A}_{\alpha\beta},\hat{B}_{\alpha\beta})$ individually satisfy the Gauss and Codazzi-Mainardi compatibility equations:
\begin{subequations}
\begin{align}
& {K}_{1212}+\text{det} [B_{\alpha \beta}] = 0,~
-{\partial}_2 {B}_{1 1}+{\partial}_1 {B}_{1 2}= 0,~
-{\partial}_2 {B}_{2 1}+{\partial}_1 {B}_{2 2}= 0; \label{gcm1} \\
& \hat{K}_{1212}+ \text{det} [\hat B_{\alpha \beta}]= 0,~
-\hat{\partial}_2 \hat{B}_{1 1}+\hat{\partial}_1 \hat{B}_{1 2}= 0,~
-\hat{\partial}_2 \hat{B}_{2 1}+\hat{\partial}_1 \hat{B}_{2 2}= 0,
\label{gcm2}
\end{align}
\label{gcm}%
\end{subequations}
where $K_{1212}$ and $\hat{K}_{1212}$ are the only independent components of the Riemannian curvature tensors $K_{\alpha\beta\mu\nu}$ and $\hat{K}_{\alpha\beta\mu\nu}$  associated with $A_{\alpha\beta}$ and $\op{A}_{\alpha\beta}$, respectively; $\partial$ and $\op{\partial}$ denote the covariant derivatives with respect to the induced Levi-Civita connections $\Gamma^\mu_{\alpha\beta}=\frac{1}{2}A^{\mu\nu}(A_{\mu\alpha,\beta}+A_{\mu\beta,\alpha}-A_{\alpha\beta,\nu})$ and $\op{\Gamma}^\mu_{\alpha\beta}=\frac{1}{2}\op{A}^{\mu\nu}(\op{A}_{\mu\alpha,\beta}+\op{A}_{\mu\beta,\alpha}-\op{A}_{\alpha\beta,\nu})$, respectively; here,  $[A^{\alpha\beta}]=[A_{\alpha\beta}]^{-1}$ and $[\op{A}^{\alpha\beta}]=[\op{A}_{\alpha\beta}]^{-1}$.
The total stretching strain and the total bending strain tensors, defined as
$\mathbb{E}=E_{\alpha\beta}\boldsymbol{A}^\alpha\otimes\boldsymbol{A}^\beta=\frac{1}{2}(\hat{A}_{\alpha\beta}-A_{\alpha\beta})\boldsymbol{A}^\alpha\otimes\boldsymbol{A}^\beta$ and $\boldsymbol{\Lambda}=\Lambda_{\alpha\beta}\boldsymbol{A}^\alpha\otimes\boldsymbol{A}^\beta=(\hat{B}_{\alpha\beta} - B_{\alpha\beta})\boldsymbol{A}^\alpha\otimes\boldsymbol{A}^\beta$, respectively, measure the relative first and second fundamental forms of the current configuration with respect to the reference configuration of the mid-surface.

\subsection{The natural configuration}
The natural (relaxed, stress/moment free) configuration, in the presence of defects, cannot be realized as a connected  isometric embedding of the mid-surface $\omega$ in $\mathbb{R}^3$~\cite{rcgupta17}. It can however be realized as an appropriate projection on $\omega$ of an isometric embedding ${\boldsymbol{\chi}}:\mathscr{B}\to\mathbb{M}^3$ within a hypothetical 3D non-Riemannian space $\mathbb{M}^3$ (the material space) equipped with a symmetric material metric and a material connection with components $g_{ij}$ and $L^k_{ij}$, respectively, with respect to the natural coordinates $\theta^i$. If $a_{\alpha\beta}$ (symmetric positive-definite) and $b_{\alpha\beta}$ (symmetric) represent the first and the second fundamental form, respectively, of the mid-surface of the shell in the natural configuration then, in accordance with the Kirchhoff-Love kinematical assumption \cite{ciar3}, the components  $g_{ij}$ can be defined as
\begin{subequations}
\begin{align}
&{g}_{\alpha\beta}= {\boldsymbol{\chi}}_{,\alpha} \cdot  {\boldsymbol{\chi}}_{,\beta} = {a}_{\alpha\beta}-2\zeta\,{b}_{\alpha\beta} +\zeta^2 \,{c}_{\alpha\beta},\\
&{g}_{\alpha 3}=g_{3\alpha}= {\boldsymbol{\chi}}_{,\alpha} \cdot  {\boldsymbol{\chi}}_{,3} =0,~\text{and}~
{g}_{33}=  {\boldsymbol{\chi}}_{,3} \cdot  {\boldsymbol{\chi}}_{,3} = 1,
\end{align}
\label{metric-natural:defn}%
\end{subequations}
where $c_{\alpha\beta}= {a}^{\mu\nu}{b}_{\mu\alpha}{b}_{\nu\beta}$ is the (symmetric) third fundamental form of the natural configuration. The metric $g_{ij}$ is positive-definite for small $\zeta$~\cite{rcgupta17}. We introduce $a=\text{det}[a_{\alpha\beta}]$, $\varepsilon_{\alpha\beta}=a^{\frac{1}{2}}e_{\alpha\beta}$,   $\varepsilon^{\alpha\beta}=a^{-\frac{1}{2}}e_{\alpha\beta}$, $g=\text{det}[g_{ij}]$,   $\varepsilon_{ijk}=g^{\frac{1}{2}}e_{ijk}$, and $\varepsilon^{ijk}=g^{-\frac{1}{2}}e_{ijk}$. Then, at $\zeta=0$, $g=a$, $\varepsilon^{\alpha\beta 3}=\varepsilon^{\alpha\beta}$, and $\varepsilon_{\alpha\beta 3}=\varepsilon_{\alpha\beta}$. Let $[a^{\alpha\beta}]=[a_{\alpha\beta}]^{-1}$ and $[g^{ij}]=[g_{ij}]^{-1}$. We use $\nabla$ to denote the covariant derivative with respect to the Levi-Civita connection $s^\mu_{\alpha\beta}=\frac{1}{2}a^{\mu\tau}(a_{\tau\beta,\alpha}+a_{\tau\alpha,\beta}-a_{\alpha\beta,\tau})$ and $\tilde\nabla$ for the covariant derivative with respect to the Levi-Civita connection $S^k_{ij}=\frac{1}{2}g^{km}(g_{mj,i}+g_{mi,j}-g_{ij,m})$. A subscript semicolon ($(\cdot)_{;k}$, etc.) is used to denote the covariant derivative with respect to the material connection $L^k_{ij}$. The pair $(a_{\alpha\beta},b_{\alpha\beta})$, unlike their counterpart in the reference and the current configurations, do not satisfy the Gauss and Codazzi-Mainardi compatibility equations. They satisfy the incompatible Gauss and Codazzi-Mainardi equations, as derived in Appendix~\ref{incompgcm}, with sources of incompatibility given in terms of defect and metric anomaly densities.

\label{natural}

\subsection{The defect densities}
\label{defectdef}

\begin{figure}[t!]
	\centering
	\includegraphics[scale=0.6]{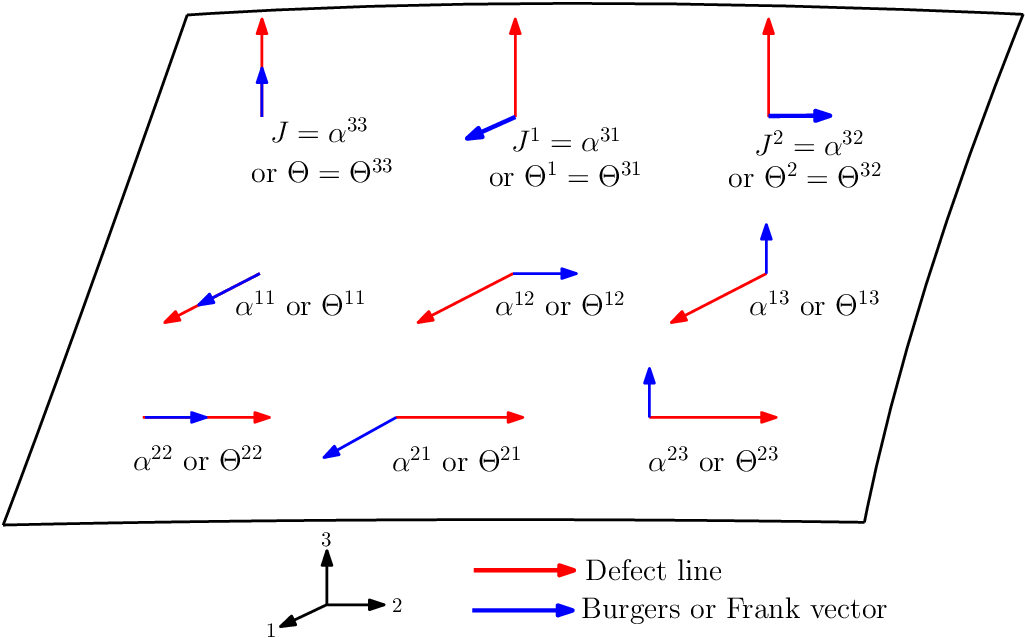}
	\caption{Individual components of dislocations and  disclinations on a Kirchhoff-Love shell.}
	\label{individual-defects}
\end{figure}

The densities of disclinations in the 3D shell are associated with  the fourth-order Riemann-Christoffel curvature tensor of the 3D material space with components $\tilde\Omega_{klj}{}^i = L^i_{lj,k}- L^i_{kj,l}+L^h_{lj} L^i_{kh}-L^h_{kj} L^i_{lh}$ \cite{rcgupta15, rcgupta17}. By definition, $\tilde\Omega_{(kl)j}{}^i=0$. We assume $\tilde{\Omega}_{ij(kl)} = 0$, where the covariant components $\tilde\Omega_{ijkl}=\tilde\Omega_{ijk}{}^m g_{ml}$, i.e., we neglect metrical disclinations \cite{rcgupta16, rcgupta17} in the shell, which otherwise give rise to the non-preservation of the inner product of tangent vectors in the material space (with respect to the material metric under parallel transport using the material connection); these are related to generalized disclinations~\cite{zhang18}. Let $\Omega_{ijkl}(\theta^\alpha)=\tilde\Omega_{ijkl}(\theta^\alpha,\zeta=0)$. Due to the skew-symmetries with respect to the first two and the last two indices of $\Omega_{ijkl}$, there are only nine independent components of the defect density. These are $\Theta^{ij}(\theta^\alpha)=\frac{1}{4}\varepsilon^{ikl}(\theta^\alpha,\zeta=0) \varepsilon^{jmn}(\theta^\alpha,\zeta=0)\Omega_{klmn}(\theta^\alpha)$.  It follows that 
\begin{equation}
\Omega_{klmn} = a e_{ikl} e_{jmn} \Theta^{ij}.
\end{equation}
The first index $i$  of $\Theta^{ij}$ denotes the direction of the defect line (i.e., the orthogonal direction to the plane of the Frank loop), while the second index $j$ denotes the direction of the associated Frank vector, see Figure~\ref{individual-defects}. The diagonal components ($i=j$) represent disclinations of the wedge type and the off-diagonal components ($i\ne j$) represent disclinations of the twist type. Two families of disclinations are distinguished based on whether their defect lines are transverse to the mid-surface $\omega$ (the Frank loops lie within $\omega$),  or whether the defect lines are tangential to the mid-surface $\omega$ (the Frank loops lie transverse to $\omega$).  The former kind of disclinations include the wedge type $\Theta(\theta^\alpha)=\Theta^{33}(\theta^\alpha)$ and the twist type $\Theta^{\rho}(\theta^\alpha)=\Theta^{3\rho}(\theta^\alpha)$. The latter kind are represented by $\Theta^{\mu k}$, out of which $\Theta^{11}$ and $\Theta^{22}$ are of the wedge type, and $\Theta^{12}$ and $\Theta^{\mu 3}$ are of the twist type. Disclinations with arbitrary orientation can be decomposed in terms of these families. The notion of a transverse, or out-of-surface, Frank loop is not realizable for monolayered shells; disclinations densities $ \Theta^{\mu k}$ can hence appear only in multilayered shells.  

The densities of dislocations in the 3D shell are associated with the torsion tensor of the 3D material space with components
$\tilde{T}_{ij}{}^k(\theta^i)=L^k_{[ij]}(\theta^i)$ \cite{rcgupta15, rcgupta17}. By definition, $\tilde{T}_{(ij)}{}^k=0$.  Let $T_{ij}{}^k(\theta^\alpha)=\tilde{T}_{ij}{}^k(\theta^\alpha,\zeta=0)$. The skew symmetry in the torsion tensor yields nine independent components given by
$\alpha^{ij}(\theta^\alpha)=\frac{1}{2}\varepsilon^{ikl}(\theta^\alpha,\zeta=0)T_{kl}{}^j(\theta^\alpha)$. As a result,
\begin{equation}
T_{12}{}^3 = \sqrt{a} \alpha^{33},~ T_{12}{}^\mu = \sqrt{a} \alpha^{3\mu}, ~\text{and}~ T_{3 \alpha}{}^k = \varepsilon_{\alpha \beta} \alpha^{\beta k}.
\end{equation} 
The index $i$ of $\alpha^{ij}$ denotes the direction of the defect line (i.e., the orthogonal direction to the plane of the Burgers parallelogram) while the index $j$ stands for the direction of the corresponding Burgers vector, see Figure~\ref{individual-defects}.  The diagonal components ($i=j$) are dislocation densities of the screw type and the off-diagonal components ($i\ne j$) are the densities of the edge type. As with disclinations, we distinguish between two families of dislocations based on whether their defect lines are transverse to the mid-surface $\omega$ (the Burgers parallelograms lie within $\omega$), or whether the defect lines are tangential to the mid-surface $\omega$ (the Burgers parallelograms lie transverse to $\omega$). 
The former kind of dislocations include the edge type $J^\mu(\theta^\alpha)=\alpha^{3\mu}(\theta^\alpha)$ and the screw type ${J}(\theta^\alpha)=\alpha^{33}(\theta^\alpha)$. 
The latter kind are represented by $\alpha^{\mu k}$, 
out of which $\alpha^{11}$ and $\alpha^{22}$ are of the screw type, and $\alpha^{12}$, $\alpha^{21}$ and  $\alpha^{\rho 3}$ are of the edge type. Dislocations with arbitrary orientation can be decomposed in terms of elements of these families. The notion of a transverse, or out-of-surface, Burgers parallelogram is not realizable for monolayered shells; dislocation densities $\alpha^{\mu k}$ can hence appear only in multilayered shells. 

The densities of the metric anomalies (point defects, growth strains, thermal strains, etc.) are represented by the non-metricity of the 3D material space with components
$\tilde{Q}_{kij}(\theta^i)=-g_{ij;k}(\theta^i)=-g_{ij,k}+L^m_{ki} g_{mj}+L^m_{kj}g_{im}$ \cite{rcgupta16, rcgupta17, rcgupta20}. 
By definition, $\tilde{Q}_{k[ij]}=0$. Let ${Q}_{kij}(\theta^\alpha)=\tilde{Q}_{kij}(\theta^\alpha,\zeta=0)$. Of the eighteen independent non-metricity components, the six components $Q_{k\alpha 3}$ measure the tilting of the normal direction, the three components $Q_{k33}$ measure the thickness distention, the six components $Q_{\mu\alpha\beta}$ represent purely in-surface metric anomalies (for instance those arising from in-plane growth strains or from point defects in 2D crystals~\cite{rcgupta20}), and the remaining three components $Q_{3\alpha\beta}$ represent the bending or curvature metric anomalies. The latter could arise due to differential growth and differential thermal deformation in multilayered shells~\cite{rcgupta20}. The densities of defects and metric anomalies, as introduced above, can not be prescribed arbitrarily but have to satisfy a system of identities; see Appendix~\ref{conslaws}.

\section{The inhomogeneous F{\"o}ppl-von K{\'a}rm{\'a}n shallow shell equations}
\label{vk}

The inhomogeneous F{\"o}ppl-von K{\'a}rm{\'a}n shallow shell equations are a pair of coupled partial differential relations for the determination of the stress function and out-of-surface deformation of the thin shell surface with sources in terms of defect and metric anomaly densities. The following derivation of these equations is based on the geometrical notions developed in the previous section, in addition to Appendix~\ref{incompgcm} and Appendix~\ref{conslaws},  under further assumptions on the magnitude of deformations, strains, and defect fields.

\subsection{The strain measures}

Let the reference mid-surface be described in a Monge representation $\boldsymbol{R}(\theta^\alpha)=\theta^\alpha\mathbf{e}_\alpha+\text{w}^0(\theta^\alpha)\mathbf{e}_3$, where $\{\mathbf{e}_i\}$ are a set of right handed Cartesian bases vectors for $\mathbb{R}^3$ and $\text{w}^0(\theta^\alpha)$ is the height function with respect to the flat surface. The mid-surface in the current configuration is described using the in-surface displacements $\text{u}^\alpha (\theta^\beta)$ and height $\text{w}(\theta^\alpha)$ (with respect to the flat surface) such that $\op{\boldsymbol{R}}(\theta^\alpha)=(\text{u}^\alpha (\theta^\beta) + \theta^\alpha) \mathbf{e}_\alpha+\text{w}(\theta^\alpha)\mathbf{e}_3$. In accordance  with the F{\"o}ppl-von K{\'a}rm{\'a}n kinematics \cite{koiter1966, NaghdiVongsarnpigoon1983}, we assume $\text{u}^\alpha$, and its derivatives, to be $O(\epsilon)$ and both $\text{w}^0$, $\text{w}$, and their derivatives, to be $O(\epsilon^{\frac{1}{2}})$. The assumption on the height functions amounts to restricting our attention to only shallow shells. Consequently, 
 $A_{\alpha\beta}=\delta_{\alpha\beta}+\text{w}^0_{,\alpha}\text{w}^0_{,\beta}+o(\epsilon)$, $B_{\alpha\beta}=\text{w}^0_{,\alpha\beta}+o(\epsilon^{\frac{1}{2}})$,  $\op{A}_{\alpha\beta}=\delta_{\alpha\beta}+ \text{u}^\gamma_{,\alpha} \delta_{\gamma \beta} + \text{u}^\gamma_{,\beta} \delta_{\gamma_\alpha} +  \text{w}_{,\alpha}\text{w}_{,\beta} +o(\epsilon)$, and $\op{B}_{\alpha\beta}=\text{w}_{,\alpha\beta}+o(\epsilon^{\frac{1}{2}})$. The components of the \textit{total} stretching and bending strain fields, given by  $E_{\alpha\beta}=\frac{1}{2}(\hat{A}_{\alpha\beta}-A_{\alpha\beta})$ and $\Lambda_{\alpha\beta}=(\hat{B}_{\alpha\beta}-B_{\alpha\beta})$, respectively, satisfy the compatibility conditions
\begin{subequations}
\begin{align}
& e^{\alpha\beta}e^{\mu\lambda} \left( E_{\alpha\mu,\beta\lambda}+B_{\alpha\mu}\Lambda_{\beta\lambda}+\frac{1}{2}\Lambda_{\alpha\mu}\Lambda_{\beta\lambda} \right) =0,\label{compatibility:fvk1}\\
& {\Lambda}_{11,2} - {\Lambda}_{12,1}= 0,~\text{and}~
{\Lambda}_{12,2} - {\Lambda}_{2 2,1}= 0\label{compatibility:fvk2}
 \end{align}
 \label{compatibility:fvk}
 \end{subequations}
to the leading order, as can be derived immediately using \eqref{gcm}.

\begin{figure}[t!]
	\centering
	\includegraphics[scale=0.6]{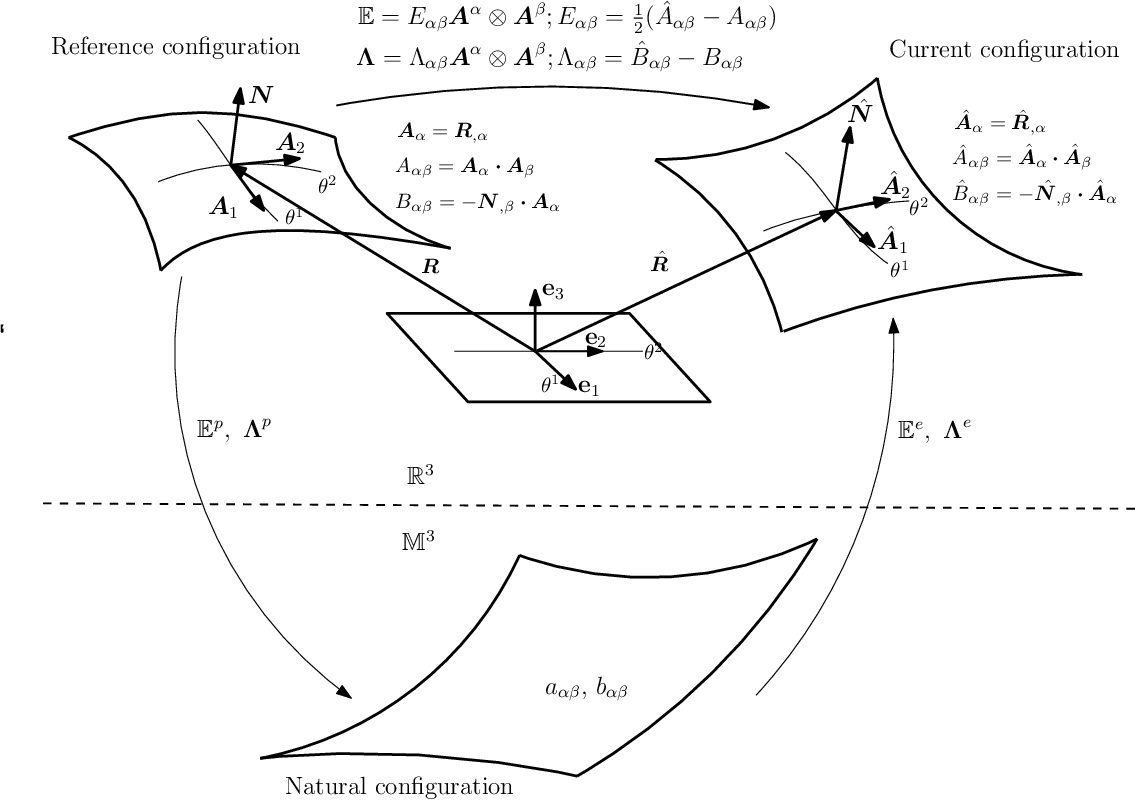}
	\caption{The reference, current, and the natural configurations of the shell mid-surface.}
	\label{decomp}
\end{figure}

The notion of \textit{elastic} stretching and bending strain is introduced as the energetic dual of the in-surface stress and bending moment fields, respectively, in the shell. The former is assumed to be symmetric and positive-definite while the latter to be symmetric. Assuming the {elastic} stretching and bending strain fields (and their derivatives) to be of order $O(\epsilon)$ and $O(\epsilon^{\frac{1}{2}})$  \cite{rcgupta17}, respectively, we write them as $\mathbb{E}^e = E^e_{\alpha\beta} \boldsymbol{A}^\alpha\otimes\boldsymbol{A}^\beta$ and $\boldsymbol{\Lambda}^e = \Lambda^e_{\alpha\beta} \boldsymbol{A}^\alpha\otimes\boldsymbol{A}^\beta$, to their leading order; here, and elsewhere, the superscript $e$ denotes {elastic} and should not be taken for an index. The basis vectors $\boldsymbol{A}^\alpha$, used in the above expressions, should be retained only to the leading order (which is $\mathbf{e}_\alpha$). The surface fundamental forms in the natural configuration, to their leading order, are taken as $a_{\alpha \beta} = \op{A}_{\alpha \beta} - 2 E^e_{\alpha\beta}$ and $b_{\alpha \beta} = \op{B}_{\alpha\beta} - \Lambda^e_{\alpha\beta}$. It then follows that $a_{\alpha \beta} = \delta_{\alpha \beta} + O(\epsilon)$, $b_{\alpha \beta} = O(\epsilon^{\frac{1}{2}})$, and $s^{\alpha}_{\beta \gamma}=O(\epsilon)$. The \textit{plastic} strain fields, $\mathbb{E}^p = E^p_{\alpha\beta} \boldsymbol{A}^\alpha\otimes\boldsymbol{A}^\beta$ and $\boldsymbol{\Lambda}^p = \Lambda^p_{\alpha\beta} \boldsymbol{A}^\alpha\otimes\boldsymbol{A}^\beta$, are introduced such that $E^p_{\alpha\beta} = \frac{1}{2}(a_{\alpha \beta} - {A}_{\alpha \beta})$ and $\Lambda^p_{\alpha\beta} = b_{\alpha \beta} - {B}_{\alpha\beta}$; here, and elsewhere, the superscript $p$ denotes {plastic} and should not be taken for an index. The total stretching strain is then additively decomposable into elastic and plastic counterparts, $\mathbb{E}=\mathbb{E}^e+\mathbb{E}^p$ with $E_{\alpha\beta}={E}^e_{\alpha\beta}+{E}^p_{\alpha\beta}$, to a leading order of $O(\epsilon)$. The total bending strain is additively decomposable into elastic and plastic counterparts, $\boldsymbol{\Lambda}=\boldsymbol{\Lambda}^e+\boldsymbol{\Lambda}^p$ with ${\Lambda}_{\alpha\beta}={\Lambda}^e_{\alpha\beta}+{\Lambda}^p_{\alpha\beta}$, to a leading order of $O(\epsilon^{\frac{1}{2}})$. The various mid-surface configurations and the  strain measures are illustrated in Figure~\ref{decomp}.

\subsection{Assumption on the order of defect density fields} \label{defectorder}

We assume that the dislocation density fields, for which both the Burgers parallelogram and the Burgers vector lie on the tangent plane of the mid-surface, are of order $O(\epsilon)$, while those for which either the Burgers parallelogram or the Burgers vector lie transverse to the tangent plane of the mid-surface are of order $O(\epsilon^{\frac{1}{2}})$. Hence, $J^\mu = O(\epsilon)$, ${J} = O(\epsilon^{\frac{1}{2}})$, and $\alpha^{\mu k} = O(\epsilon^{\frac{1}{2}})$. Similarly we assume that the disclination density fields, for which both the Frank loop and the Frank vector lie on the tangent plane of the mid-surface, are of order $O(\epsilon)$, whereas those for which either the Frank loop or the Frank vectors lie transverse to the tangent plane of the mid-surface are of order $O(\epsilon^{\frac{1}{2}})$. Hence, $\Theta= O(\epsilon)$, $\Theta^{\mu}= O(\epsilon^{\frac{1}{2}})$, and $\Theta^{\mu k}= O(\epsilon^{\frac{1}{2}})$. Finally, we assume the extensional metric anomalies to be of order $O(\epsilon)$ and the  curvature metric anomalies to be of order $O(\epsilon^{\frac{1}{2}})$, i.e., $Q_{\mu\alpha\beta}= O(\epsilon)$ and  $Q_{3\alpha\beta}= O(\epsilon^{\frac{1}{2}})$. 
Accordingly, identity \eqref{irrot-1} implies that there exist symmetric functions $q^0_{\alpha\beta}$, $q'_{\alpha\beta}$, and $q''_{\alpha\beta}$, of order $O(\epsilon)$, $O(\epsilon^{\frac{1}{2}})$, and $O(\epsilon)$, respectively, such that \cite{rcgupta20}
\begin{equation}
 \tilde{q}_{\alpha\beta}(\theta^\alpha,\zeta)=q^0_{\alpha\beta}+\zeta q'_{\alpha\beta}+\zeta^2 q''_{\alpha\beta}~\text{and}~\tilde{q}_{i3}(\theta^\alpha,\zeta)=0.
\label{bprel3-cons1}
\end{equation}
Further terms in the Taylor series expansion of $\tilde{q}_{\alpha\beta}$ (with respect to $\zeta$) are inconsequential to our discussion. We have $Q_{\mu\alpha\beta}=-2q^0_{\alpha\beta,\mu}$ and $Q_{3\alpha\beta}=-2q'_{\alpha\beta}$. The components $Q_{ij3}=0$ are either zero or order $O(\epsilon^{\frac{3}{2}})$. 
The derivatives of the defect density and metric anomaly fields are assumed to respect the order of the defect field.

\subsection{The strain incompatibility relations}

The elastic and plastic strains are \text{incompatible} in the sense that they do not satisfy relations of the kind given in \eqref{compatibility:fvk}. Due to this, they are not related to deformation fields in a way that the total strains are. The incompatibility of strains is due to the presence of defect and metric anomaly densities over the surface. The strain incompatibility relations are obtained by reducing the relations between the Riemannian and the non-Riemannian structures of the theory to their leading order using the assumptions prescribed above. Following the derivation in Appendix~\ref{incompgcm}, we obtain the the required \textit{strain incompatibility relations} as
\begin{subequations}
	\begin{align}
	& e^{\alpha\beta}e^{\mu\nu} \left(E^p_{\alpha\mu,\beta\nu} - q^0_{\alpha\mu,\beta\nu} \right) + \text{det}[B_{\alpha \beta} -  \hat\Lambda^p_{\alpha \beta}] - \frac{1}{2} \left[ \text{w}^0 , \text{w}^0 \right] =\Theta-2e_{\alpha \beta} J^\alpha_{,\beta}  - (J)^2,	\label{s-i-r-fvk-1} \\
	& \hat\Lambda^p_{11,2}- \hat\Lambda^p_{1 2,1}=-\Theta^2 + J_{,1}, ~\text{and} \label{s-i-r-fvk-12}\\
	&  \hat\Lambda^p_{12,2}- \hat\Lambda^p_{2 2,1}=\Theta^1 + J_{,2},\label{s-i-r-fvk-22}
	\end{align}
	\label{s-i-r-fvk}%
\end{subequations}
where we have used the identity $\text{det}[B_{\alpha \beta}] = \frac{1}{2}\left[ \text{w}^0 , \text{w}^0 \right] $ and the substitutions
\begin{equation}
 \hat\Lambda^p_{11} =-\Lambda^p_{11} - q'_{11} - 2\alpha^{21},~ \hat\Lambda^p_{22} = -\Lambda^p_{22} - q'_{22} + 2\alpha^{12}, ~\text{and}~ \hat\Lambda^p_{12} = \hat\Lambda^p_{21}  = -\Lambda^p_{12} - q'_{12} - \alpha^{22} + \alpha^{11}.
\end{equation}
Equations \eqref{s-i-r-fvk} are the required strain incompatibility equations for incompatible strains $E^p_{\alpha \beta}$ and $\Lambda^p_{\alpha \beta}$ with sources of incompatibility given in terms of dislocations ($J$, $J^\mu$, $\alpha^{\mu \nu}$), disclinations ($\Theta$, $\Theta^\mu$), and metric anomalies ($q^0_{\alpha \beta}$, $q'_{\alpha \beta}$), all defined over the surface. The strain incompatibility equations  \eqref{s-i-r-fvk} have not appeared earlier in the scientific literature, to the best of our knowledge. The incompatibility equations will be combined with the equations of equilibrium, discussed next, to pose boundary value problems for the determination of stress field and out-of-plane shell deflection in response to a given prescription of defects/non-metricity.   We note that dislocation densities $\alpha^{\mu 3}$ and disclination densities $\Theta^{\alpha \beta}$, both of which are related to each other in \eqref{disl1213disc12}, do not contribute towards the strain incompatibility. The disclination densities $\Theta^{\alpha 3}$, although not present explicitly in \eqref{s-i-r-fvk}, are related to $\Theta^\mu$ and $\alpha^{\mu \beta}$ through \eqref{bprel21-fvk22}. Whenever the defect/non-metricity fields are nilpotent, i.e., they lead to vanishing incompatibilities, Equations \eqref{s-i-r-fvk} are reduced to compatibility equations of the form \eqref{compatibility:fvk}.


\subsection{The governing equations}
The equilibrium equations for a F{\"o}ppl-von K{\'a}rm{\'a}n shell are written in terms of the stress components $ \sigma^{\alpha\beta}$ and moment components $m^{\alpha\beta}$, both symmetric, as  \cite{koiter1966}
\begin{subequations}
 \begin{align}
& \sigma^{\alpha\beta}{}_{,\beta}=0~\text{and}\label{equib:fvk1}\\
& m^{\alpha\beta}{}_{,\alpha\beta} -(B_{\alpha\beta}+\Lambda_{\alpha\beta})\sigma^{\alpha\beta} =0. \label{equib:fvk2}
 \end{align}
\label{equib:fvk}%
 \end{subequations}
The stress and moment components are assumed to be related to elastic stretching and bending strains through the isotropic, materially uniform,  linear elastic constitutive relations \cite{koiter1966}
\begin{subequations}
	\begin{align}
	&  \sigma^{\alpha\beta}=\frac{E}{1-\nu^2}\left( (1-\nu)\delta^{\alpha\mu}\delta^{\beta\nu}+\nu  \delta^{\alpha \beta} \delta^{\mu\nu}\right)E^e_{\mu\nu}~\text{and}\label{consti1}\\
	&  m^{\alpha\beta}=D\left( (1-\nu)\delta^{\alpha\mu}\delta^{\beta\nu}+\nu  \delta^{\alpha \beta} \delta^{\mu\nu}\right)\Lambda^e_{\mu\nu},	\label{consti2}
	\end{align}
\end{subequations}
where $E$ and $D$ are stretching and bending modulus for the 2D surface, and $\nu$ is the Poisson's ratio. The equilibrium equation  \eqref{equib:fvk1} is identically satisfied if the stress components are expressed in terms of the scalar Airy stress function $\Phi$ such that $\sigma^{\alpha\beta}=e^{\alpha\mu}e^{\beta\nu}\Phi_{,\mu\nu}$. We consider the strain compatibility equation \eqref{compatibility:fvk1}, use $E_{\alpha\beta} = E^e_{\alpha \beta} + E^p_{\alpha \beta}$, and substitute $E^e_{\alpha \beta}$ in terms of stress (and therefore the stress function) using \eqref{consti1} to  obtain the \textit{first F{\"o}ppl-von K{\'a}rm{\'a}n equation} as
\begin{equation}
\Delta^2\Phi+ \frac{E}{2} [\text{w},\text{w}] = -E\left(\lambda^p-\frac{1}{2} \left[\text{w}^0,\text{w}^0\right]\right), \label{Firstvk}
\end{equation}
with
\begin{equation}
\lambda^p=e^{\alpha\beta}e^{\mu\nu}E^p_{\alpha\mu,\beta\nu}.
\end{equation}
On the other hand, we use the constitutive equation \eqref{consti2} and ${\Lambda}^e_{\alpha\beta} = {\Lambda}_{\alpha\beta} - {\Lambda}^p_{\alpha\beta}$ in \eqref{equib:fvk2}, while replacing stress in terms of the Airy stress function, to derive the \textit{second F{\"o}ppl-von K{\'a}rm{\'a}n equation} as
\begin{equation}
D\Delta^2\text{w}-[\text{w},\Phi]  =-D  \left(\Omega^p-\Delta^2\text{w}^0\right), \label{Secvk}
\end{equation}
with
\begin{equation}
\Omega^p= -\nu \Lambda^p_{\alpha\alpha,\beta\beta}-(1-\nu)\Lambda^p_{\alpha\beta,\alpha\beta}.
\end{equation}
The two F{\"o}ppl-von K{\'a}rm{\'a}n equations \eqref{Firstvk} and \eqref{Secvk} are used to determine the out-of-plane deformation $\text{w}$ and stress $\sigma^{\alpha \beta}$ in the shell, with a known reference shape $\text{w}^0$, for a prescription of defect and metric anomaly densities. The presence of defects and metric anomalies, contained in $\lambda^p$ and $\Omega^p$, will be brought out clearly in the governing equations presented in the following sections. The analogous F{\"o}ppl-von K{\'a}rm{\'a}n equations for a plate are recovered by imposing $\text{w}^0 = 0$.

We digress briefly to consider the case when strains $E^p_{\alpha \beta}$ and $\Lambda^p_{\alpha\beta}$ are compatible, i.e., they satisfy equations of the type \eqref{compatibility:fvk}. This can happen when the given defect/non-metricity distribution is nilpotent. Then there will exist a scalar field $u(\theta^\alpha)$ such that $\Lambda^p_{\alpha\beta} = u_{,\alpha \beta}$. The equations \eqref{Firstvk} and \eqref{Secvk} subsequently take the form
\begin{subequations}
\begin{align}
& \Delta^2\Phi+ \frac{E}{2} [\text{w},\text{w}] = \frac{E}{2} \left[(\text{w}^0 +u),(\text{w}^0 + u)\right]~\text{and} \\
& D\Delta^2\text{w}-[\text{w},\Phi]  =D  \Delta^2(\text{w}^0+u),
\end{align}
\end{subequations}
respectively. The energetically optimal solution (in the sense that it leads to minimum energy) of these equations is $\Phi=0$ and $\text{w} = \text{w}^0 + u$, assuming that the boundary conditions are also satisfied. The assumption will be violated if $\text{w}$, or its gradient, is prescribed over the boundary; we ignore this possibility in writing our solution. The solution is in fact such that it yields $E^e_{\alpha \beta} = 0$ and $\Lambda^e_{\alpha\beta}=0$ and hence a vanishing strain energy.

\subsection{Recovering known results}

Starting with the general results derived above, we can recover some known results already available in the literature. We consider
only in-surface densities of defects and metric anomalies to be present, i.e., let only 
 densities $\Theta$, $J^\alpha$, and $q^0_{\alpha\beta}$ to be non-zero. We also assume $\Lambda^p_{\alpha\beta}=0$ for it has no source in the considered densities and also so that there is no indeterminacy in the model. We obtain $\Omega^p=0$ and 
$\lambda^p  = \Theta-2 e_{\alpha \beta} J^\alpha_{,\beta} + e^{\alpha\beta}e^{\mu\nu} q^0_{\alpha\mu,\beta\nu}$.
Substituting these in the F{\"o}ppl-von K{\'a}rm{\'a}n equations \eqref{Firstvk} and \eqref{Secvk} we get
\begin{subequations}
\begin{align}
& \Delta^2\Phi+ \frac{E}{2} [\text{w},\text{w}] = \frac{E}{2} \left[\text{w}^0,\text{w}^0\right] - E(\Theta-2 e_{\alpha \beta} J^\alpha_{,\beta} + e^{\alpha\beta}e^{\mu\nu} q^0_{\alpha\mu,\beta\nu})~\text{and} \\
& D\Delta^2\text{w}-[\text{w},\Phi]  =D  \Delta^2\text{w}^0.
\end{align}
\end{subequations}
On further assuming isotropic non-metricity, i.e., $q^0_{\alpha\mu} = Q\delta_{\alpha \mu}$, where $Q(\theta^\alpha)$ is a scalar field, and considering the reference surface to be flat ($\text{w}^0=0$) we obtain
\begin{equation}
\Delta^2\Phi+ \frac{E}{2} [\text{w},\text{w}] = - E(\Theta-2 e_{\alpha \beta} J^\alpha_{,\beta} + \Delta Q)~\text{and}~ D\Delta^2\text{w}-[\text{w},\Phi]  =0.
\end{equation}
These equations appear in the book by Nelson \cite[pp. 217–238]{nelson-book} although with defect densities written in terms of Dirac measures (which are used to represent isolated singularities), with $Q$ denoting a density of point defects. The other scenario, where results are previously known, is when only metric anomalies are considered (no dislocations and disclinations). This case was discussed in detail in our recent work \cite{rcgupta20}, where the derived formalism was connected with the existing work on morphology of growing thin shells  \cite{liang-mahadevan11} and thermal deformations for F{\"o}ppl-von K{\'a}rm{\'a}n shells. In the absence of dislocations and disclinations, $\lambda^p  =e^{\alpha\beta}e^{\mu\nu} q^0_{\alpha\mu,\beta\nu}$ and 
$\Omega^p= \nu q'_{\alpha\alpha,\beta\beta}+(1-\nu)q'_{\alpha\beta,\alpha\beta}$, where, in the context of growth, $q^0_{\alpha\mu}$ and $q'_{\alpha\beta}$ are to be interpreted as extensional and bending growth strains \cite{rcgupta20}.

\section{Dislocations in F{\"o}ppl-von K{\'a}rm{\'a}n plates} \label{dislocations}

Let disclinations and metric anomalies be absent. Therefore, $\Theta =0$, $\Theta^\mu =0$, $\Theta^{\mu k} = 0$, $q^0_{\alpha \beta}=0$, and $q'_{\alpha \beta}=0$. The identities \eqref{dislconslaws}$_1$ and \eqref{disl1213disc12} then imply that $\alpha^{\rho \lambda}{}_{,\rho} =0$ and $\alpha^{\nu 3}$ are constant, respectively.  Also, let the reference surface be flat, i.e., $\text{w}^0=0$, thereby restricting the discussion to plates. The strain incompatibility equations \eqref{s-i-r-fvk} are reduced to
\begin{subequations}
	\begin{align}
	& e^{\alpha\beta}e^{\mu\nu} E^p_{\alpha\mu,\beta\nu} + \text{det}[\hat\Lambda^p_{\alpha \beta}] =-2e_{\alpha \beta} J^\alpha_{,\beta}  - (J)^2,	\label{s-i-r-fvk-onlyd-1} \\
	& \hat\Lambda^p_{11,2}- \hat\Lambda^p_{1 2,1}=J_{,1}, ~\text{and}~\hat\Lambda^p_{12,2}- \hat\Lambda^p_{2 2,1}=J_{,2}, \label{s-i-r-fvk-onlyd-2}
	\end{align}
	\label{s-i-r-fvk-onlyd}%
\end{subequations}
where $\hat\Lambda^p_{11} =-\Lambda^p_{11} - 2\alpha^{21}$, $\hat\Lambda^p_{22} = -\Lambda^p_{22} + 2\alpha^{12}$, and  $\hat\Lambda^p_{12} = -\Lambda^p_{12} - \alpha^{22} + \alpha^{11}$. The solution for \eqref{s-i-r-fvk-onlyd-2} is of the form $\hat\Lambda^p_{\alpha \beta} = -e_{\alpha \beta} J + \Upsilon_{\alpha,\beta}$, where $\Upsilon_{\alpha}(\theta^\beta)$ are components of a vector field such that, to ensure the symmetry of $\Lambda^p_{\alpha \beta}$, $e_{\alpha \beta} \Upsilon_{\alpha,\beta} = 2J$. We can resolve the indeterminacy in $\Upsilon_{\alpha}$, and hence in $\hat\Lambda^p_{\alpha \beta}$, by positing existence of a scalar field $\psi (\theta^\alpha)$ such that $\Upsilon_{\alpha} = e_{\alpha\beta}\psi_{,\beta}$ and $\psi=0$ on the boundary. Therefore $\hat\Lambda^p_{\alpha \beta} = -e_{\alpha \beta} J + e_{\alpha\gamma}\psi_{,\gamma\beta}$ and the components of plastic bending strain can be written as 
\begin{equation}
\Lambda^p_{11} = -2\alpha^{21} - \psi_{,12},~\Lambda^p_{22} = 2\alpha^{12} + \psi_{,12},~\text{and}~\Lambda^p_{12} = - \alpha^{22} + \alpha^{11} + J - \psi_{,22} \label{plastbenstronlyd}
\end{equation}
such that
\begin{equation}
\Delta \psi = 2J. \label{laplscrew}
\end{equation}
Substituting  $\hat\Lambda^p_{\alpha \beta}$ into \eqref{s-i-r-fvk-onlyd-1} yields
\begin{equation}
e^{\alpha\beta}e^{\mu\nu} E^p_{\alpha\mu,\beta\nu}  + \frac{1}{2} [\psi,\psi]=-2e_{\alpha \beta} J^\alpha_{,\beta}. 
\end{equation} 
This can be used in \eqref{Firstvk} to obtain the first F{\"o}ppl-von K{\'a}rm{\'a}n equation with dislocations:
\begin{equation}
\frac{1}{E}\Delta^2\Phi+ \frac{1}{2} [\text{w},\text{w}] = 2e_{\alpha \beta} J^\alpha_{,\beta} +  \frac{1}{2} [\psi,\psi]. \label{Firstvk-onlyd}
\end{equation}
The second F{\"o}ppl-von K{\'a}rm{\'a}n equation with dislocations is obtained by substituting $\Omega^p$, as solved in terms of $\Lambda^p_{\alpha \beta}$ given in \eqref{plastbenstronlyd}, into \eqref{Secvk} as
\begin{equation}
D\Delta^2\text{w}-[\text{w},\Phi]  =2 D \Delta\left(\alpha^{12} - \alpha^{21}\right), \label{Secvk-onlyd}
\end{equation}
where we have also used the conservation laws~\eqref{bprel21-fvk22}. Equations \eqref{laplscrew}, \eqref{Firstvk-onlyd}, and \eqref{Secvk-onlyd} are the required governing equations in order to determine the deformation $\text{w}$ and stress in the plate. Here, recall that, $J^\alpha$ and $J$ are edge and screw dislocation densities, respectively, with dislocation lines threading normal to the 2D surface. The dislocations corresponding to densities $\alpha^{\mu \nu}$, on the other hand, have their lines and Burgers vectors both tangential to the 2D surface. The densities $\alpha^{\mu 3}$, which represent dislocations with Burgers vector orthogonal to the surface and defect lines tangential to the surface, play no role whatsoever. The densities $\alpha^{\mu \nu}$ influence the equations only through their skew part; they do appear completely in the calculation of moments $m^{\alpha \beta}$ in the plate. If the screw dislocation density $J$ is vanishingly small then $\Delta \psi =0$ (with $\psi=0$ over the boundary) implies that $\psi(\theta^\alpha)=0$ over the plate. In such a case \eqref{Firstvk-onlyd}, without the last term, and \eqref{Secvk-onlyd} suffice as the governing equations. 

\subsection{Isolated dislocations} \label{isoldesl}
Whereas isolated edge dislocations $J^\mu$ and screw dislocations $J$ appear as a point on the surface (the respective dislocation lines are orthogonal to the surface), isolated dislocations represented by $\alpha^{\mu \nu}$ can appear as a line on the surface. The former are denoted in terms of a Dirac measure supported on a point in $\mathbb{R}^2$ and the later in terms of a Dirac measure supported on a line in $\mathbb{R}^2$. For instance if we consider an isolated edge dislocation whose line direction is along the $\theta^2 = 0$ line and whose Burgers vector is along the $\theta^1 =0$ line, then $\alpha^{12} =b \delta_L$, where $b$ is the magnitude of the Burgers vector, $L$ represents the $\theta^2=0$ line, and $\int_{\mathbb{R}^2} \phi \delta_L dA = \int_L \phi dL$ for every smooth scalar field $\phi$ compactly supported on $\mathbb{R}^2$. The conservation law $\alpha^{12}{}_{,1} = 0$ is trivially satisfied. The governing equations for the determination of stress and transverse deformation field due to such a defect are
\begin{equation}
	\frac{1}{E}\Delta^2\Phi+ \frac{1}{2} [\text{w},\text{w}] =0 ~\text{and}~D\Delta^2\text{w}-[\text{w},\Phi]  =2D b \Delta\delta_L. \label{fvk-insurd1}
\end{equation}
These equations are identical to the classical F{\"o}ppl-von K{\'a}rm{\'a}n equations for a plate with a distributed transverse loading of the form $2Db\Delta\delta_L$. In the inextensional limit ($E\rightarrow \infty$), \eqref{fvk-insurd1}$_1$ reduces to $[\text{w},\text{w}] =0$, which implies a vanishing Gaussian curvature of the deformed surface. An isolated edge dislocation, of the type discussed above, has been dealt previously in the context of linearized plate theory~\cite{nabarrokostlan78}.  On the other hand, if there is an isolated edge dislocation with line in the direction orthogonal to the plate surface and Burgers vector along the $\theta^2 =0$ line then $J^1 = b \delta_o$, where $\delta_o$ is a Dirac measure supported on a point (here the origin $\theta^1=\theta^2=0$). The governing equations are
\begin{equation}
	\frac{1}{E}\Delta^2\Phi+ \frac{1}{2} [\text{w},\text{w}] =2b(\delta_o)_{,2} ~\text{and}~D\Delta^2\text{w}-[\text{w},\Phi]  =0. \label{fvk-insurd2}
\end{equation}
This problem has been studied earlier by Seung and Nelson \cite{SeungNelson88}. In the inextensible limit, the defect acts as a source for the Gaussian curvature of the deformed plate. Finally, we consider an isolated screw dislocation whose line direction and Burgers vector both are along the direction orthogonal to the surface, i.e., $J = b \delta_o$.  The governing equations, now involving \eqref{laplscrew}, are
\begin{equation}
	\frac{1}{E}\Delta^2\Phi+ \frac{1}{2} [\text{w},\text{w}] =\frac{1}{2} [\psi,\psi],~D\Delta^2\text{w}-[\text{w},\Phi]  =0,~\text{and}~\Delta \psi = 2b\delta_o. \label{fvk-insurd3}
\end{equation}
The scalar field $\psi$ can be constructed using the Green's function for the 2D Poisson's equation such that it vanishes on the boundary of the plate. In the inextensible limit $[\text{w},\text{w}] =[\psi,\psi]$; the Gaussian curvature is hence determined from the potential $\psi$. The problem of an isolated screw dislocation threading normally through the plate surface has been previously discussed only in the context of linearized plate theory~\cite{eshelbystroh51}. We note that a flat solution (i.e., $\text{w} = 0$) can possibly be a solution to both \eqref{fvk-insurd2} and \eqref{fvk-insurd3}, where the stress function will necessarily satisfy $\Delta^2\Phi =-2Eb(\delta_o)_{,1}$ and $2\Delta^2\Phi =E [\psi,\psi]$, respectively. However, a globally flat solution is inadmissible for \eqref{fvk-insurd1} as it violates \eqref{fvk-insurd1}$_2$. 

An interesting scenario arises when the isolated dislocation threads the plate surface at an inclination~\cite{saitoetal72}. For definiteness we consider a screw dislocation passing through the surface at the origin with the defect line inclined along $\cos \theta \mathbf{e}_1 + \sin \theta \mathbf{e}_3$. When $\theta = 90^\circ$ we recover the case of screw dislocation discussed above. In the present situation, the inclined isolated defect can be decomposed into four components: screw components $\alpha^{11} = b\delta_o \cos^2\theta$ and $J=b\delta_o \sin^2\theta$, and edge components $\alpha^{13}=\alpha^{31} = b\delta_o \cos\theta \sin \theta$.  The component $\alpha^{11}$ however needs to satisfy the conservations law $\alpha^{11}{}_{,1} + \alpha^{21}{}_{,2} = 0$, which requires a non-zero $\alpha^{21}$. Therefore, the inclined screw dislocation can be considered only if we additionally allow for an edge dislocation of the type $\alpha^{21}$ such that the conservation law is satisfied. The governing equations will then have source terms from $J$ and $\alpha^{21}$.

\subsection{Numerical examples} \label{disexp}

\begin{figure}[t!]
\begin{subfigure}{.248\textwidth}
  \centering
\includegraphics[scale=0.4]{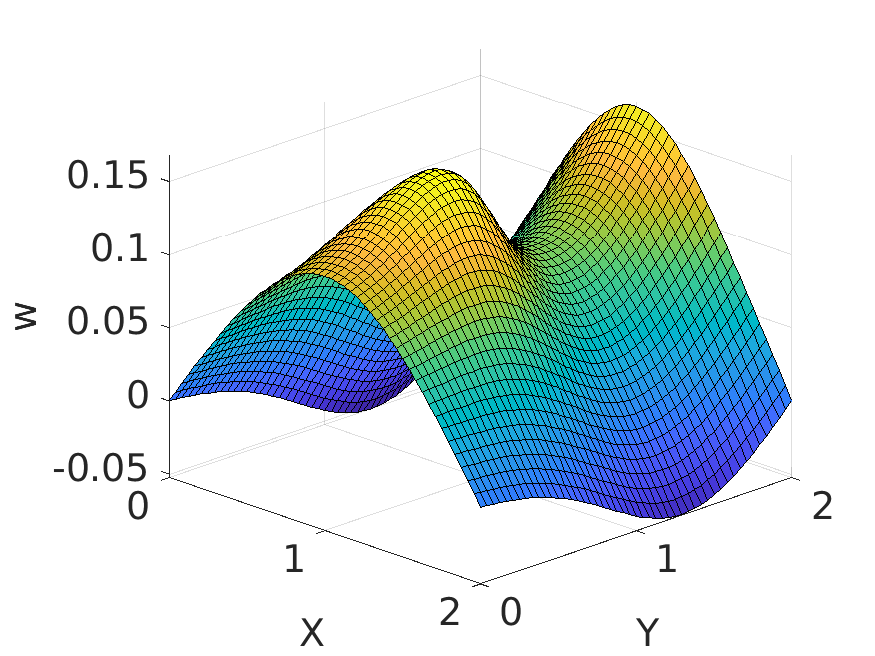}
\end{subfigure}%
\hspace{15pt}
\begin{subfigure}{.744\textwidth}
  \centering
\includegraphics[scale=0.7, trim = 0 66 35 71, clip]{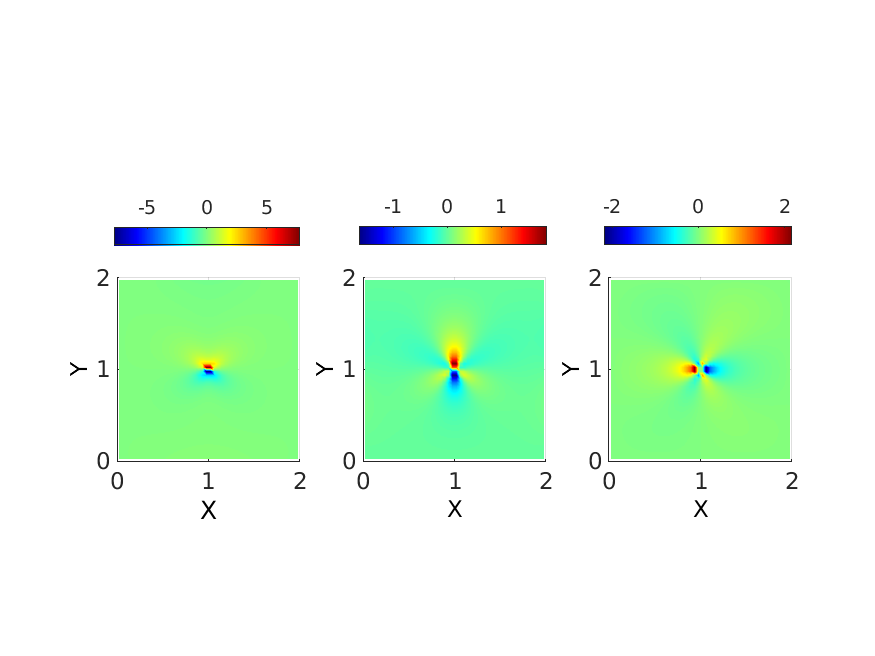}
\end{subfigure}%
\caption{The deformation $\text{w}$ and stress fields ($\sigma^{11}$, $\sigma^{22}$, and $\sigma^{12}$) for an isolated edge dislocation with defect line piercing normally through the plate at its centre.}
\label{edge1}
\end{figure}

We will now provide solutions to different problems involving various dislocation configurations on a square shaped plate with free boundary conditions (as given in \eqref{bc}). The solution methodology depends on a finite element framework which we have implemented based on a mixed variational principle as discussed in Appendix~\ref{numerics}. We state our results in terms of arbitrarily prescribed length (l) and force (f) units. The side length of the square plate, the deformation $\text{w}$, and the magnitude of Burgers vector $b$, all have units as l.  The constitutive parameters $E$ and $D$ have units of l$^{-1}$f and lf, respectively. The stresses have units of l$^{-1}$f. In all our simulations, we fix the size of the plate as $2 \times 2$ and take $E = 40$, $D=0.01$. The coordinate axes for the plate are represented as X (horizontal) and Y (vertical) with origin at one corner of the plate.

We first consider an isolated edge dislocation whose defect line pierces the plate normally at its centre and whose Burgers vector is along the horizontal axis of the plate. The boundary value problem includes the governing equations \eqref{fvk-insurd2} and the boundary conditions \eqref{bc}. For the fixed size of the plate, and given $E,D$, there is a critical magnitude of $b$, calculated around $0.02$, below which the plate remains flat ($\text{w}=0$)~\cite{SeungNelson88}. We take $b=0.025$ and obtain a buckled solution as a stable equilibrium solution. The deformation and the stresses are given in Figure~\ref{edge1}; $\text{w}$ is restrained at three corners of the plate so as to avoid rigid body motions. The results are in agreement with those obtained previously~\cite{SeungNelson88}. In particular, the stress values are all concentrated around the defect. It should be noted that the earlier works, for simulation purposes, took the edge dislocation as a disclination dipole and used a network model to get the desired results. Our finite element methodology is more general in that it is amenable to modelling different boundary conditions and allows for a continuous distribution of defects. 

\begin{figure}[t!]
\begin{subfigure}{.248\textwidth}
  \centering
\includegraphics[scale=0.4]{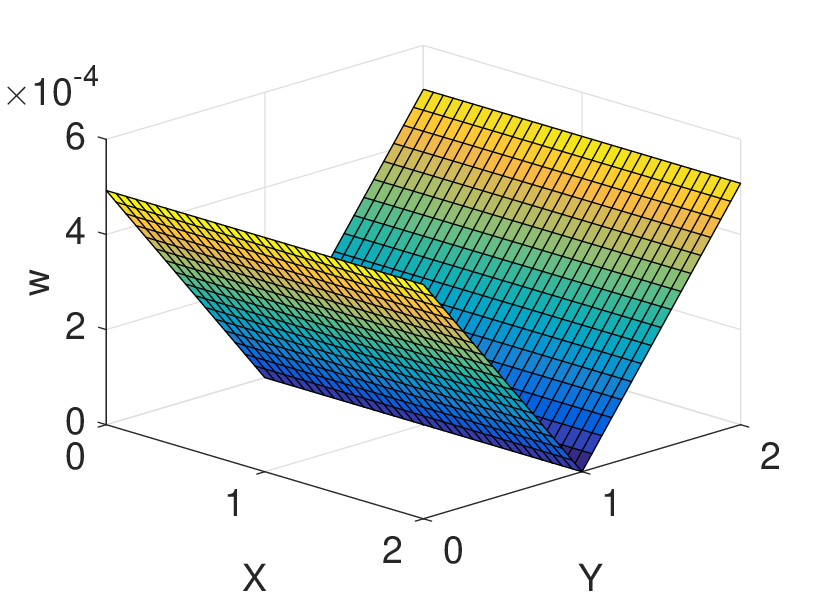}
\end{subfigure}%
\hspace{15pt}
\begin{subfigure}{.744\textwidth}
  \centering
\includegraphics[scale=0.7, trim = 0 66 35 71, clip]{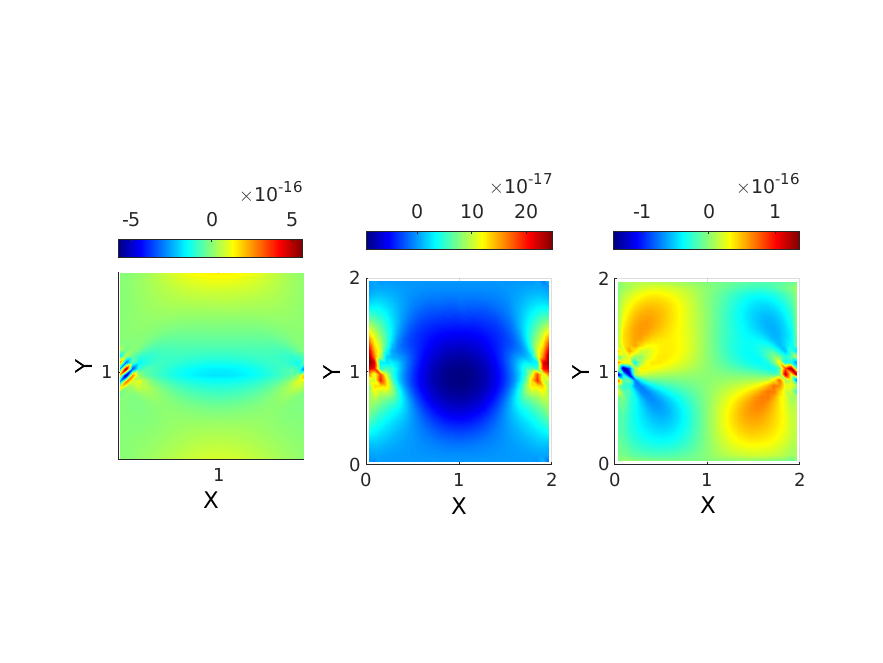}
\end{subfigure}%
\caption{The deformation $\text{w}$ and stress fields ($\sigma^{11}$, $\sigma^{22}$, and $\sigma^{12}$) for an isolated straight edge dislocation with the defect line along Y$=1$ line.}
\label{edge2}
\end{figure}

\begin{figure}[t]
\begin{subfigure}{.248\textwidth}
  \centering
\includegraphics[scale=0.4]{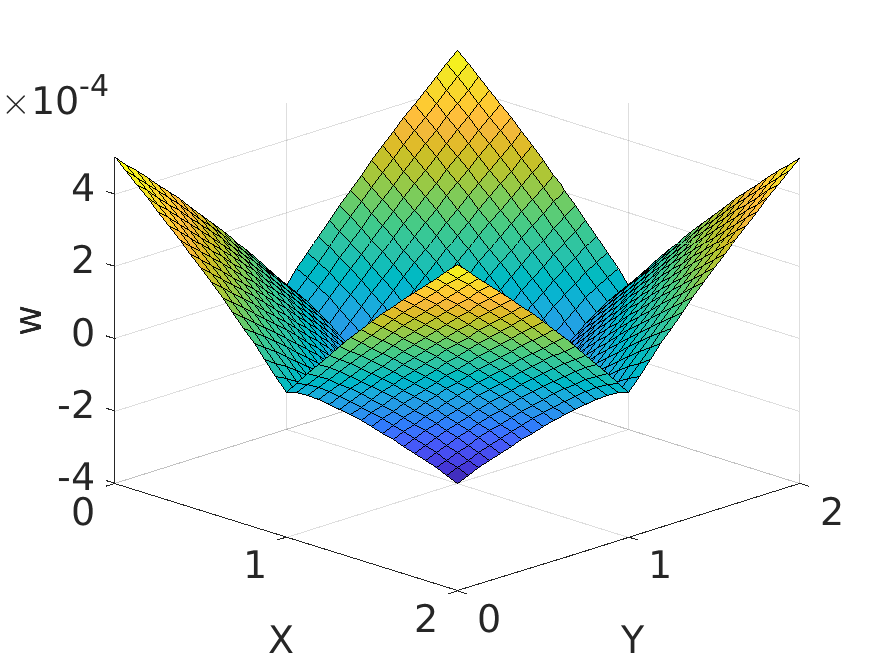}
\end{subfigure}%
\hspace{15pt}
\begin{subfigure}{.744\textwidth}
  \centering
\includegraphics[scale=0.7, trim = 0 66 35 71, clip]{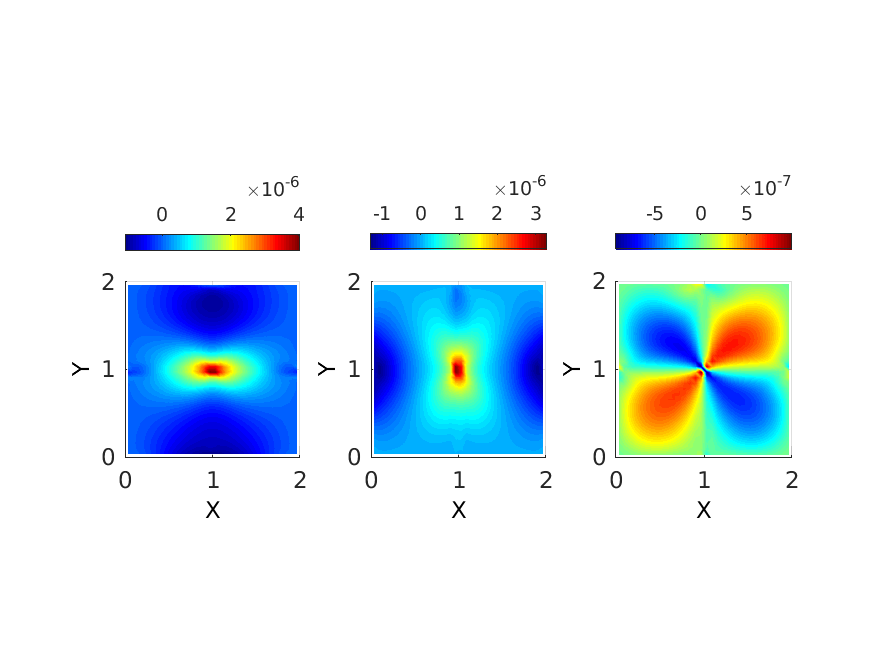}
\end{subfigure}%
\caption{The deformation $\text{w}$ and stress fields ($\sigma^{11}$, $\sigma^{22}$, and $\sigma^{12}$) for two isolated straight edge dislocations with the defect lines along Y$=1$ line and X$=1$ line.}
\label{edge3}
\end{figure}

\begin{figure}[t]
\begin{subfigure}{.248\textwidth}
  \centering
\includegraphics[scale=0.37]{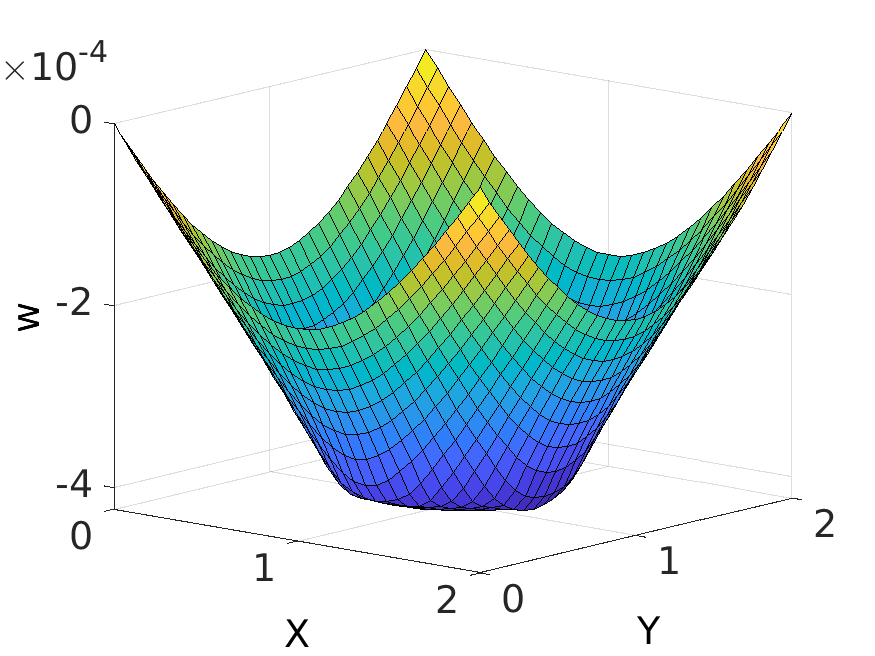}
\end{subfigure}%
\hspace{20pt}
\begin{subfigure}{.744\textwidth}
  \centering
\includegraphics[scale=0.7, trim = 30 66 35 71, clip]{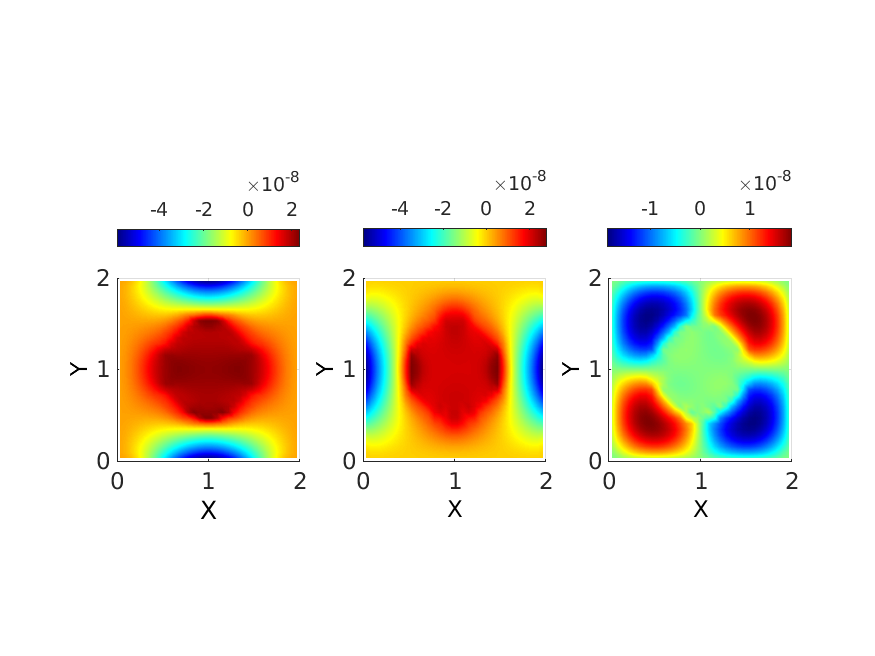}
\end{subfigure}%
\caption{The deformation $\text{w}$ and stress fields ($\sigma^{11}$, $\sigma^{22}$, and $\sigma^{12}$) for an isolated dislocation circular loop of unit diameter lying within the plate.}
\label{edge4}
\end{figure}

We next consider various configurations of isolated edge dislocations whose defect line as well as the Burgers vector lie within the plate. We discuss three cases. In each of these we have taken $2b=0.001$. First, we look at an isolated straight edge dislocation with defect line along the Y$=1$ line and Burgers vector in the in-plane orthogonal direction. Therefore the only non-trivial dislocation density component is $\alpha^{12} = b\delta_L$, as introduced in the beginning of Section~\ref{isoldesl}, where $L$ is now the Y$=1$ line. The boundary value problem includes the governing equations \eqref{fvk-insurd1} and the boundary conditions \eqref{bc}. The results are given in Figure~\ref{edge2}. Both $\text{w}$ and its normal gradient are fixed to a vanishing value at the two ends of the defect line. This is required for the implementation of the Dirac concentration over the line. The plate deforms by folding about the dislocation line with slope linearly proportional to $b$. The value of the slope is independent of variations in $E$ and $D$. These features are in agreement with the classical solution~\cite{eshelby79, nabarrokostlan78}. The stress magnitudes are negligible. In the second case, we add (to the previous case) another in-plane straight edge dislocation whose defect line is along the X$=1$ line such that the dislocation is represented by the component $\alpha^{21} = - b \delta_{L'}$, where $L'$ stands for the X$=1$ line. The governing equations are now given as
\begin{equation}
	\frac{1}{E}\Delta^2\Phi+ \frac{1}{2} [\text{w},\text{w}] =0 ~\text{and}~D\Delta^2\text{w}-[\text{w},\Phi]  =2D b \left(\Delta\delta_L + \Delta \delta_{L'}\right) \label{fvk-insurd4}
\end{equation}
supplemented with the boundary conditions. The numerical solution is given in Figure~\ref{edge3}. Both $\text{w}$ and its normal gradient are fixed to a vanishing value at the ends of the two defect lines. The deformation pattern again involves a bending about the defect lines resulting into a petal-like formation. The intersecting defect lines introduces a point of singularity at the centre. The stress values remain small but are now concentrated around the singularity. In the third case, we consider a circular dislocation loop of unit diameter around the centre of the plate. In terms of polar coordinates $(r,\theta)$, with the origin at the centre, and the corresponding basis vectors $\{\textbf{e}_r, \textbf{e}_\theta\}$, we have the dislocation density tensor $\boldsymbol{\alpha} =b \delta_C \textbf{e}_\theta \otimes \textbf{e}_r$, where $C$ is the circular loop of unit diameter around the centre and $\delta_C$ is the Dirac measure supported on the loop $C$. Using the substitutions $\textbf{e}_r = \cos \theta \textbf{e}_1 + \sin \theta \textbf{e}_2$ and $\textbf{e}_\theta = \sin \theta \textbf{e}_1 - \cos \theta \textbf{e}_2$, we can derive $\alpha^{12} = b \delta_C \sin^2 \theta$ and $\alpha^{21} = -b \delta_C \cos^2 \theta$ yielding the governing equations
\begin{equation}
	\frac{1}{E}\Delta^2\Phi+ \frac{1}{2} [\text{w},\text{w}] =0 ~\text{and}~D\Delta^2\text{w}-[\text{w},\Phi]  =2D b \Delta\delta_C. \label{fvk-insurd5}
\end{equation}
The simulation results are given in Figure~\ref{edge4}. The deformation $\text{w}$ is fixed at three corners of the plate to avoid rigid body motions. The plate bends about the circular loop while remaining flat inside the loop. The stress magnitudes are smaller than the previous case (almost by two orders) but the stress fields are distributed more non-locally. All the three cases demonstrate the rich variety of shapes which can be obtained by suitably placing the in-surface edge dislocations without generating high stresses and without any external intervention (in terms of displacements and forces).

\section{Disclinations coupled with growth in F{\"o}ppl-von K{\'a}rm{\'a}n plates} \label{disclinations}
Let dislocations be absent and the reference configuration be flat ($\text{w}^0=0$). Hence, $J^\mu=0$, $J=0$, $\alpha^{\mu j}=0$. Furthermore we identify $q^0_{\alpha \beta}$ and $q'_{\alpha \beta}$ with isotropic stretching and bending growth strains, respectively, thereby writing $q^0_{\alpha \beta}= Q \delta_{\alpha \beta}$ and $q'_{\alpha \beta}= P \delta_{\alpha \beta}$, where $Q(\theta^\alpha)$ and $P(\theta^\alpha)$ are the transverse and longitudinal components, respectively, of isotropic growth \cite{rcgupta20}. Analogously, in the context of thermal strains, we can interpret $Q$ and $P$ as the first and second order temperature fields, respectively \cite{rcgupta20}. In the absence of dislocations, identities \eqref{dislconslaws}$_1$ and \eqref{disl1213disc12} imply that $\Theta^{ij}$ is symmetric, $\Theta^{\alpha \beta} = 0$, $\Theta^1 = \Theta^{13} =  \vartheta_{,2}$, and $\Theta^2 = \Theta^{23} =  -\vartheta_{,1}$, where $\vartheta(\theta^\alpha)$ is a scalar field. The non-trivial disclination densities are therefore given in terms of two scalar fields $\Theta$ and $\vartheta$. Recall that $\Theta$ represents the density of wedge disclinations whose defect line direction and Frank vector remain both orthogonal to the surface; $\Theta^\mu$ represent twist disclinations with line direction orthogonal to the surface and Frank vector remaining tangential to the surface; $\Theta^{\mu3}$ represent twist disclinations with line direction tangential to the surface and Frank vector orthogonal to the surface. The strain incompatibility equations \eqref{s-i-r-fvk} are reduced to
\begin{subequations}
	\begin{align}
	& e^{\alpha\beta}e^{\mu\nu} E^p_{\alpha\mu,\beta\nu} + \text{det}[\hat\Lambda^p_{\alpha \beta}] = \Theta + \Delta Q,	\label{s-i-r-fvk-onlydi-1} \\
	& \hat\Lambda^p_{11,2}- \hat\Lambda^p_{1 2,1}=\vartheta_{,1}, ~\text{and}~\hat\Lambda^p_{12,2}- \hat\Lambda^p_{2 2,1}=\vartheta_{,2}, \label{s-i-r-fvk-onlydi-2}
	\end{align}
	\label{s-i-r-fvk-onlydi}%
\end{subequations}
where $\hat\Lambda^p_{11} =-\Lambda^p_{11} - P$, $\hat\Lambda^p_{22} = -\Lambda^p_{22} -P$, and  $\hat\Lambda^p_{12} = -\Lambda^p_{12}$. Following the arguments presented in Section \ref{dislocations}, we can write the solution for \eqref{s-i-r-fvk-onlydi-2} as $\hat\Lambda^p_{\alpha \beta} = -e_{\alpha \beta} \vartheta + e_{\alpha \gamma} \psi_{,\gamma\beta}$ where $\psi$ is a scalar field satisfying the Poisson's equation
 \begin{equation}
\Delta \psi = 2\vartheta, \label{lapldiscl}
\end{equation}
with $\psi=0$ on the boundary of the domain. The components of plastic bending strain can henceforth be written as 
\begin{equation}
\Lambda^p_{11} = -P - \psi_{,12},~\Lambda^p_{22} = -P + \psi_{,12},~\text{and}~\Lambda^p_{12} =    \vartheta - \psi_{,22}. \label{plastbenstrdiscl}
\end{equation}
Substituting these into \eqref{s-i-r-fvk-onlydi-1}, and combining the resulting relation with \eqref{Firstvk}, we obtain the first F{\"o}ppl-von K{\'a}rm{\'a}n equation with disclinations and isotropic growth as
\begin{equation}
\frac{1}{E}\Delta^2\Phi+ \frac{1}{2} [\text{w},\text{w}] = \frac{1}{2} [\psi,\psi] -  (\vartheta)^2 - \Theta - \Delta Q, \label{Firstvk-discl}
\end{equation}
where we can use \eqref{lapldiscl} to alternatively write the term $\frac{1}{2} [\psi,\psi] -  (\vartheta)^2$ as $-(\psi_{,12})^2 - \frac{1}{4}(\psi_{,11} - \psi_{,22})^2$.
The second F{\"o}ppl-von K{\'a}rm{\'a}n equation with disclinations and isotropic growth is obtained by substituting $\Omega^p$ into \eqref{Secvk}. For $\Lambda^p_{\alpha \beta}$ given in \eqref{plastbenstrdiscl}, we obtain $\Omega^p=(1+\nu)\Delta P$ reducing \eqref{Secvk} to
\begin{equation}
D\Delta^2\text{w}-[\text{w},\Phi]  =-D(1+\nu)\Delta P. \label{Secvk-discl}
\end{equation}
Interestingly for an auxetic plate, with $\nu = -1$, the isotropic bending growth strains will play no role whatsoever. Equations \eqref{lapldiscl}, \eqref{Firstvk-discl}, and \eqref{Secvk-discl} are the required governing equations in order to determine the deformation $\text{w}$ and stress in the plate for a given distribution of disclinations and isotropic growth strains.  We observe that, in the inextensional limit, disclinations $\Theta^\mu = \Theta^{\mu 3}$, and $\Theta$, and the in-surface growth $Q$, all appear as sources of Gaussian curvature of the deformed plate shape; the growth strain can therefore completely or partially screen the effect of disclinations. The bending growth strain field $P$, through \eqref{Secvk-discl}, provides a transverse load-like distribution over the plate. If disclinations of the type $\Theta^\mu$ (and hence $\Theta^{\mu 3}$) are absent then both $\vartheta$ and $\psi$ can be taken to be identically zero without loss of generality. 
The governing equations are then reduced to the pair
\begin{equation}
\frac{1}{E}\Delta^2\Phi+ \frac{1}{2} [\text{w},\text{w}] = - \Theta- \Delta Q~\text{and}~ D\Delta^2\text{w}-[\text{w},\Phi]  =-D(1+\nu)\Delta P. \label{Secvk-discl-sp1}
\end{equation}
Clearly, whenever growth strains $Q$, $P$ and disclination density $\Theta$ are such that $\Delta Q = -\Theta$ and $\Delta P = 0$ there is no stress and deformation response in the plate. Given $\Theta$, the ensuing Poisson's equation and Laplace equation can be solved (with appropriate boundary conditions) to obtain the growth fields which will yield no mechanical response in the plate.

Let the growth strains be absent. The governing equations for studying the mechanics of an isolated wedge disclination threading normally through the plate surface (at origin) are given by substituting $\Theta = \Omega \delta_o$, where $\Omega$ is the strength of the disclination,  and $Q=0$, $P=0$ into \eqref{Secvk-discl-sp1}. The solution to the resulting system of equations has been discussed in detail by Seung and Nelson \cite{SeungNelson88}.
On the other hand, if disclinations of the type $\Theta^1$ are present, but $\Theta^2=0$, then by the symmetry restriction, disclinations $\Theta^{13}$ of equal magnitude should also be present. In this case, $\vartheta$ is such that $\vartheta_{,1}=0$ and $\vartheta_{,2} = \Theta^1$. Hence $\Theta^1$ (and $\Theta^{13}$) is necessarily a function of only $\theta^2$ coordinate. Consequently we cannot take disclination of the form $\Theta^1=\Theta^{13} = \Omega \delta_o$, with no other defect present. We can however consider $\Theta^1=\Theta^{13} = \Omega \delta_L$ where $L$ is the $\theta^2=0$ curve, for instance. This corresponds to one twist disclination of the kind $\Theta^{13}$ (with defect line along the $\theta^2=0$ curve) and an array of twist disclinations $\Theta^1$ (with defect lines normal to the surface) distributed uniformly along the $\theta^2=0$ curve.

\subsection{Numerical examples}
\label{disclexp}

The coupling between disclinations and growth strains is now illustrated through some simple numerical simulations. As with dislocations, the system of our interest is a square shaped plate with free boundary conditions (given in \eqref{bc}); the solutions are obtained using a finite element methodology as outlined in Appendix~\ref{numerics}. The notation for units is as discussed in Section~\ref{disexp}. In all the simulations, we fix the size of the plate as $2 \times 2$ and take $E = 40$, $D=0.01$. The coordinate axes for the plate are represented as X (horizontal) and Y (vertical) with origin at one corner of the plate.  We consider isolated wedge disclinations and isotopic growth strains such that $\Theta = \pm (\pi/3) \delta_o$, $\Delta Q = \pm \pi/3$, and $\Delta P = \pm \pi/3$. We study the coupling by comparing the deformation profiles. The deformation $\text{w}$ is kept fixed (as $0$) at four corners of the plate to avoid rigid body motions and for maintaining a four-fold symmetry in the deformation pattern (particularly for cases with $\Delta P \neq 0$).

\begin{figure}[t!]
\captionsetup[subfigure]{justification=centering, font=footnotesize,aboveskip=-1pt,belowskip=-1pt}
\begin{subfigure}{.248\textwidth}
  \centering
\includegraphics[scale=0.32]{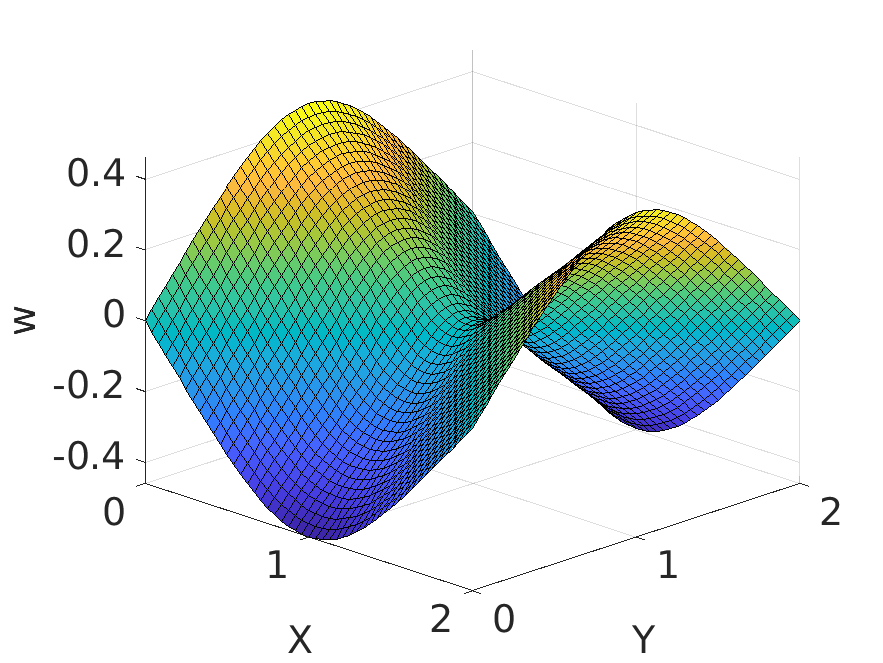}
\subcaption{$s=1,\Delta Q =0, \Delta P =0$}
\end{subfigure}%
\hspace{35pt}
\begin{subfigure}{.248\textwidth}
  \centering
\includegraphics[scale=0.32]{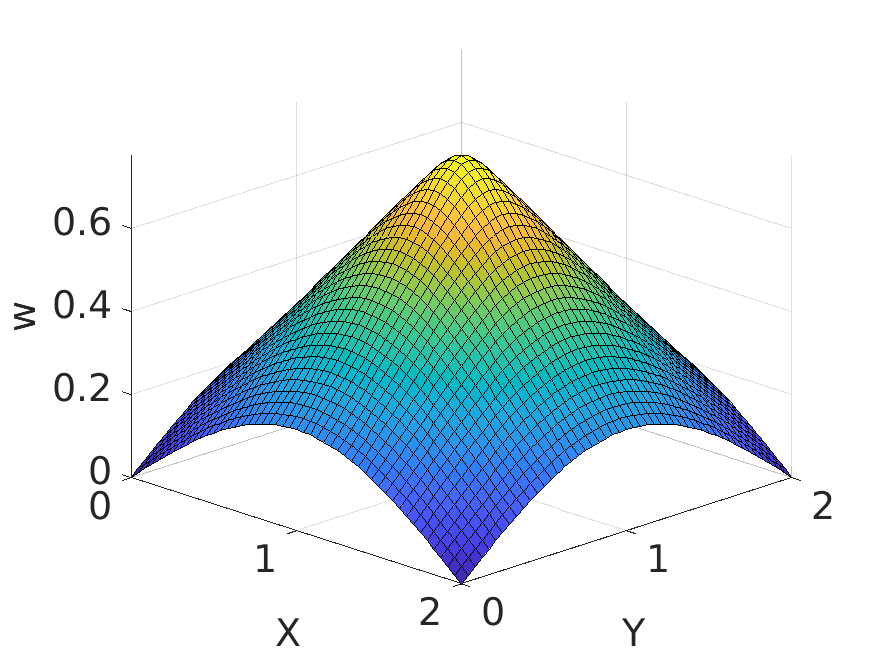}
\subcaption{$s=-1,\Delta Q =0, \Delta P =0$}
\end{subfigure}%
\hspace{35pt}
\begin{subfigure}{.248\textwidth}
  \centering
\includegraphics[scale=0.32]{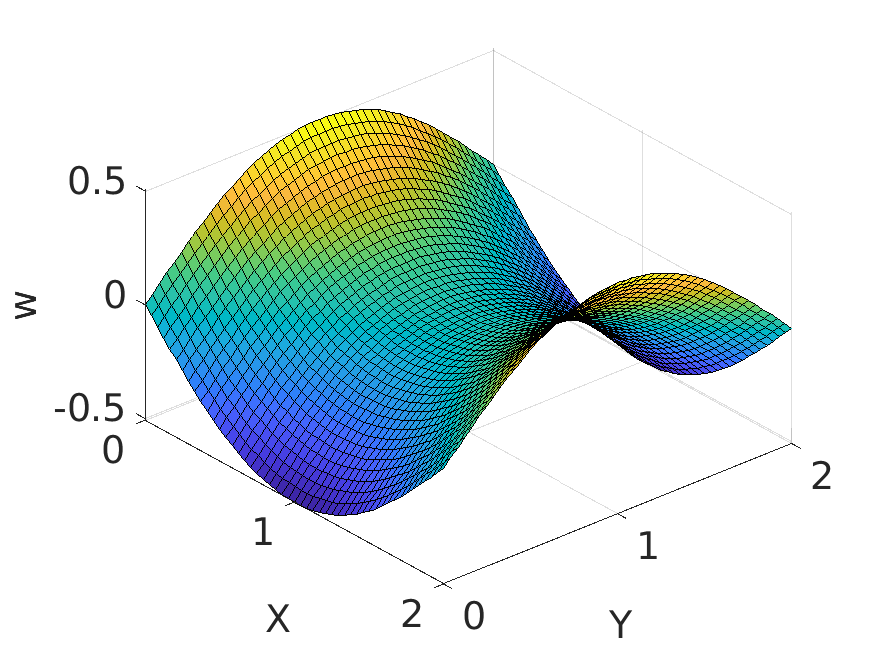}
\subcaption{$s=0,\Delta Q =\frac{\pi}{3}, \Delta P =0$}
\end{subfigure}%

\begin{subfigure}{.248\textwidth}
  \centering
\includegraphics[scale=0.32]{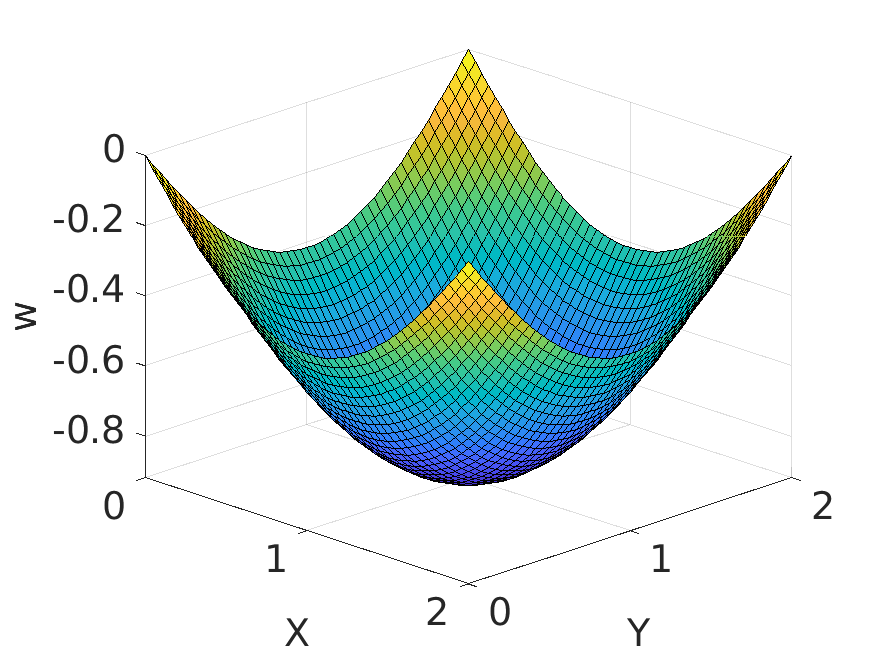}
\subcaption{$s=0,\Delta Q =-\frac{\pi}{3}, \Delta P =0$}
\end{subfigure}%
\hspace{35pt}
\begin{subfigure}{.248\textwidth}
  \centering
\includegraphics[scale=0.32]{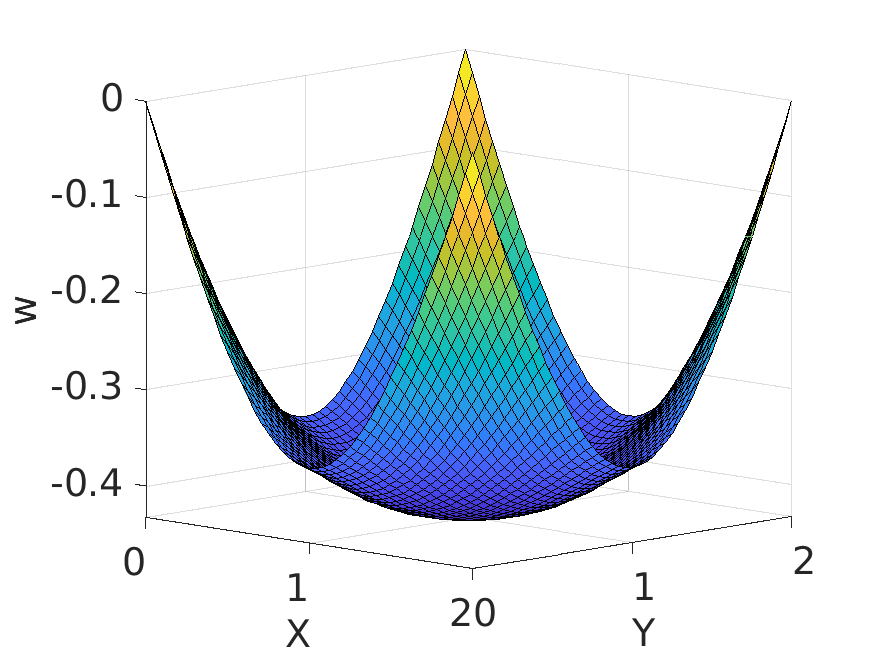}
\subcaption{$s=0,\Delta Q =0, \Delta P =\frac{\pi}{3}$}
\end{subfigure}%
\hspace{35pt}
\begin{subfigure}{.248\textwidth}
  \centering
\includegraphics[scale=0.32]{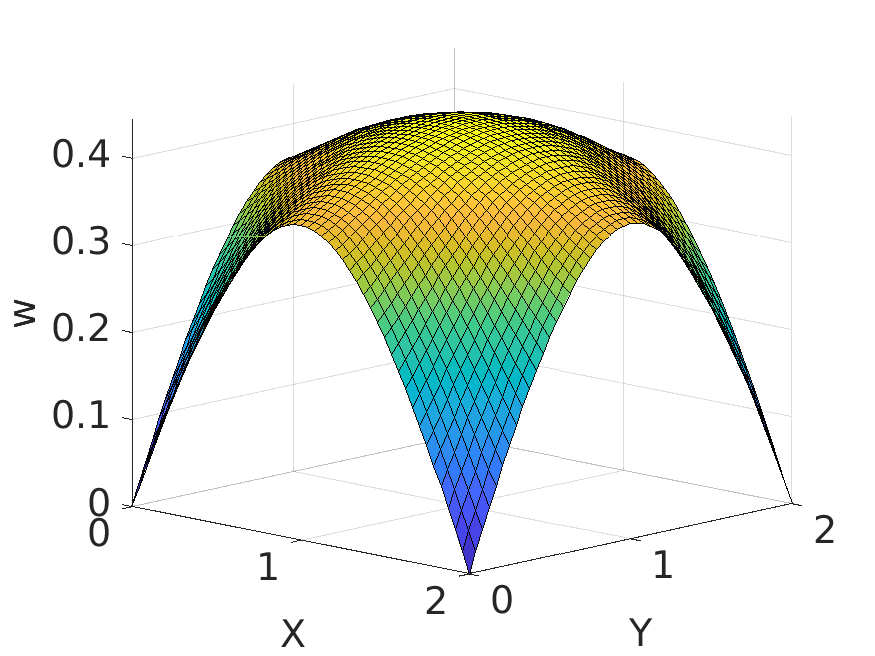}
\subcaption{$s=0,\Delta Q =0, \Delta P =-\frac{\pi}{3}$}
\end{subfigure}%
\caption{The deformation $\text{w}$ for various individual cases of isolated wedge disclinations and isotropic growth. The disclination density is of the form $\Theta=s\frac{\pi}{3}\delta_o$.}
\label{disclgrowth1}
\end{figure}

To begin with, we plot the deformations $\text{w}$ obtained individually from each of the sources, see Figure~\ref{disclgrowth1}. The saddle shape, resulting from a single negative wedge disclination ($\Theta = (\pi/3) \delta_o$, Figure~\ref{disclgrowth1}(a)), and the conical shape, resulting from a single positive wedge disclination ($\Theta =- (\pi/3) \delta_o$, Figure~\ref{disclgrowth1}(b)), are well known \cite{SeungNelson88}. The actual deformations are a regularized saddle and a regularized cone due to the contribution from elastic stretching in the first term of \eqref{Secvk-discl-sp1}$_1$ in addition to small corrections due to the boundary~\cite{Pandey21}. Recall that perfect saddle and perfect cone shapes are the result of a negative and a positive Dirac concentration in Gaussian curvature, respectively. The plots with $\Delta P = \pm \pi/3$ are those that will also appear due to a distribution of uniform transverse load of magnitude $-D(1+\nu)\Delta P$, see Equation~\eqref{Secvk-discl-sp1}$_2$, while keeping the corners of the plate fixed. The deformations with $\Delta Q = \pm \pi/3$ correspond approximately to surfaces with constant Gaussian curvature $-\Delta Q$. We write approximately because of the deviations due to extensional elasticity and boundary effects. Note that the deformation profiles, given for cases where $\Delta P =0$, are equally likely to exist with $\text{w}$ replaced by $-\text{w}$, i.e., Figures~\ref{disclgrowth1}(a-d) could have been given equivalently with shapes reflected about the XY-plane. This symmetry in solutions is evident from the boundary value problem \eqref{Secvk-discl-sp1} and \eqref{bc}, where, with $\Delta P =0$, a change in the sign of $\text{w}$ does not effect the system of equations. This is not true whenever $\Delta P \neq0$. Another distinction between the two cases comes from the fact that the results with $\Delta P = 0$ are necessarily the buckled solutions; there always exist corresponding solutions, where $\text{w}=0$, which are unstable in the considered parametric regime. On the other hand, there cannot be any solution with $\text{w}=0$ whenever $\Delta P \neq 0$.
We also mention, without providing explicit results, that stress fields for the isolated disclination cases are localized around the defect but otherwise distributed all over the plate for isotropic growth fields. 

\begin{figure}[t!]
\captionsetup[subfigure]{justification=centering, font=scriptsize,aboveskip=-1pt,belowskip=-1pt}
\begin{subfigure}{.24\linewidth}
  \centering
\includegraphics[scale=0.3]{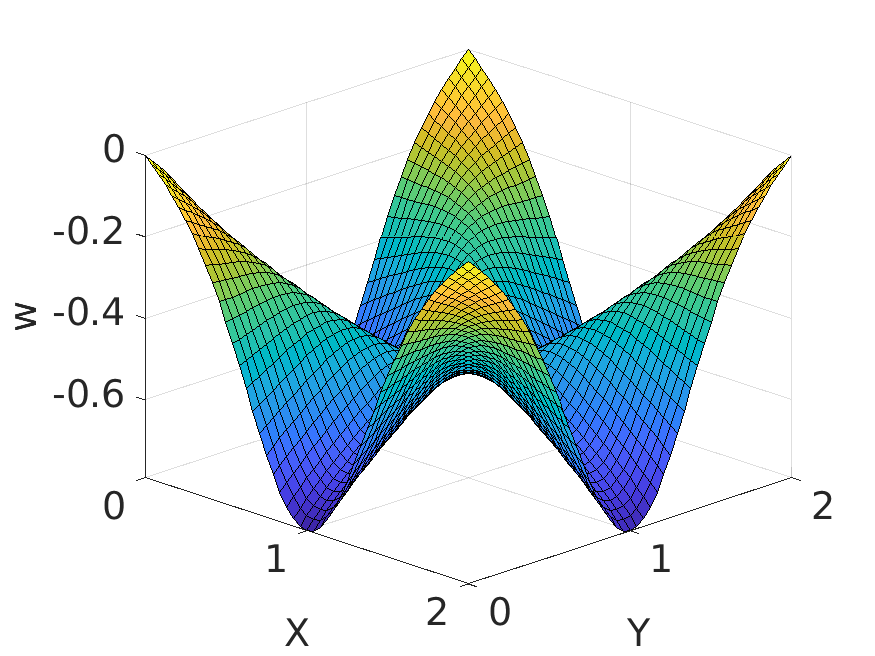}
\subcaption{$s=1,\Delta Q =\frac{\pi}{3}, \Delta P =0$}
\end{subfigure}%
\hspace{1pt}
\begin{subfigure}{.24\linewidth}
  \centering
\includegraphics[scale=0.3]{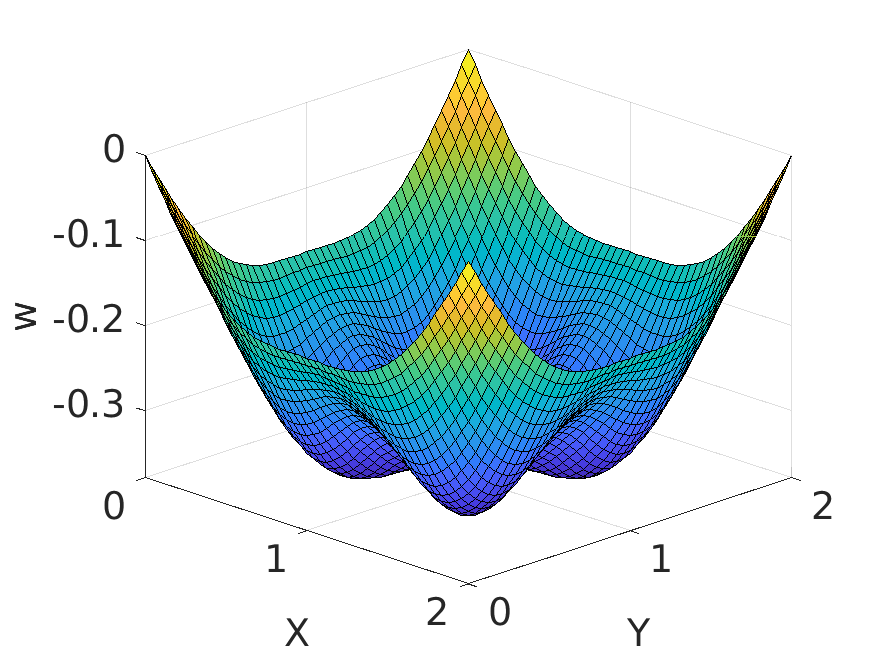}
\subcaption{$s=1,\Delta Q =-\frac{\pi}{3}, \Delta P =0$}
\end{subfigure}%
\hspace{1pt}
\begin{subfigure}{.24\linewidth}
  \centering
\includegraphics[scale=0.3]{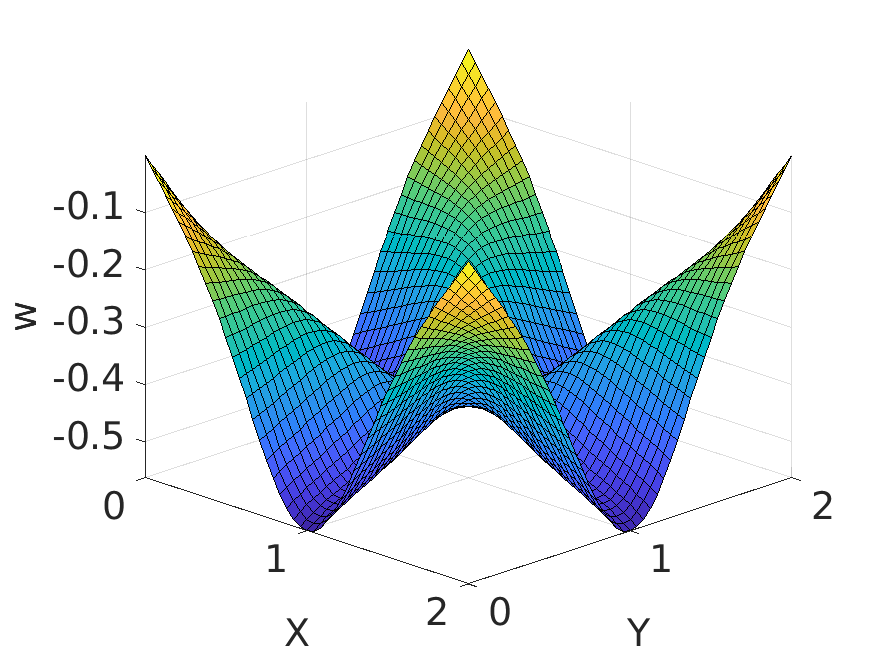}
\subcaption{$s=1,\Delta Q =0, \Delta P =\frac{\pi}{3}$}
\end{subfigure}%
\hspace{1pt}
\begin{subfigure}{.24\linewidth}
  \centering
\includegraphics[scale=0.3]{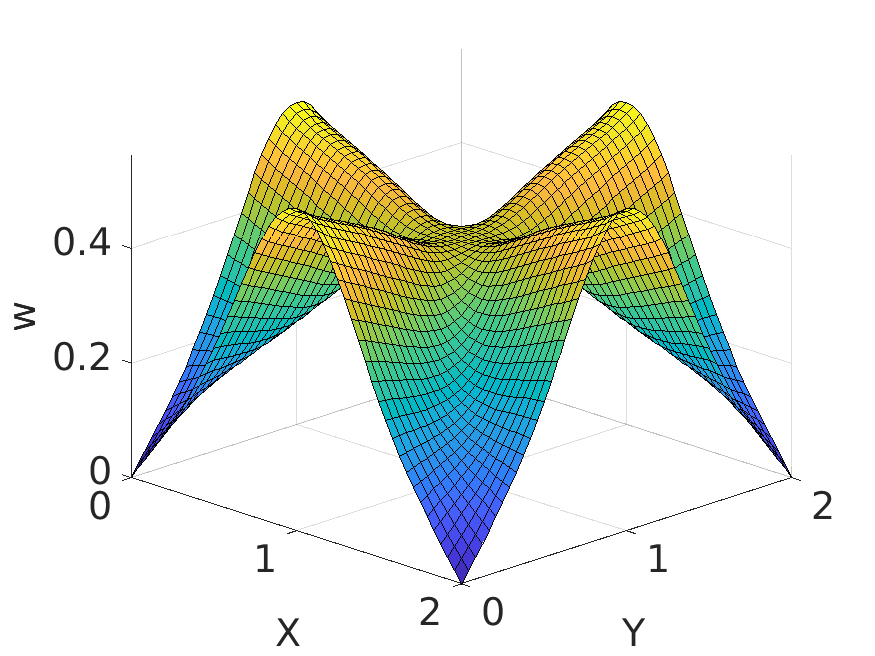}
\subcaption{$s=1,\Delta Q =0, \Delta P =-\frac{\pi}{3}$}
\end{subfigure}%

\begin{subfigure}{.24\linewidth}
  \centering
\includegraphics[scale=0.3]{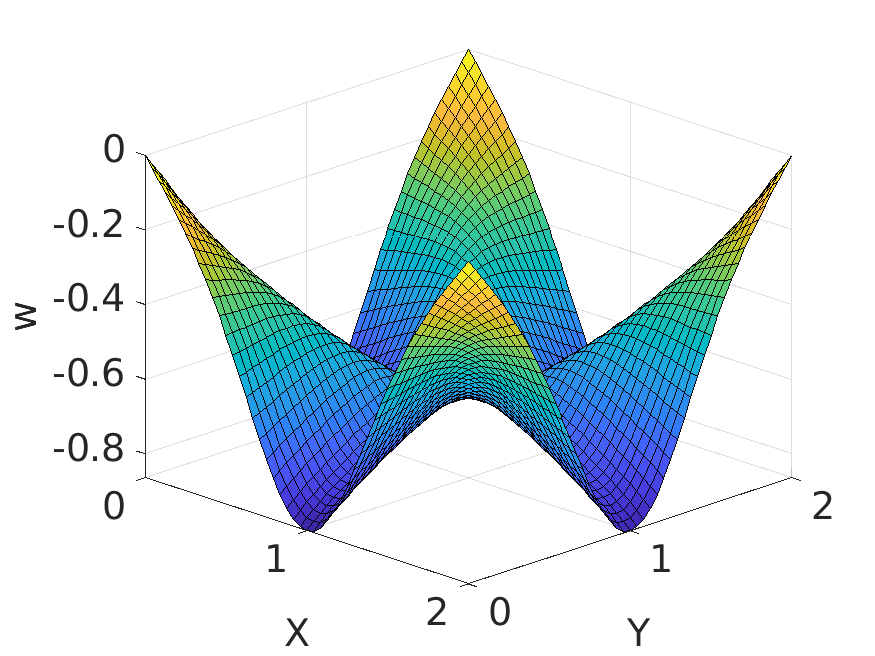}
\subcaption{$s=1,\Delta Q =\Delta P =\frac{\pi}{3}$}
\end{subfigure}%
\hspace{1pt}
\begin{subfigure}{.24\linewidth}
  \centering
\includegraphics[scale=0.3]{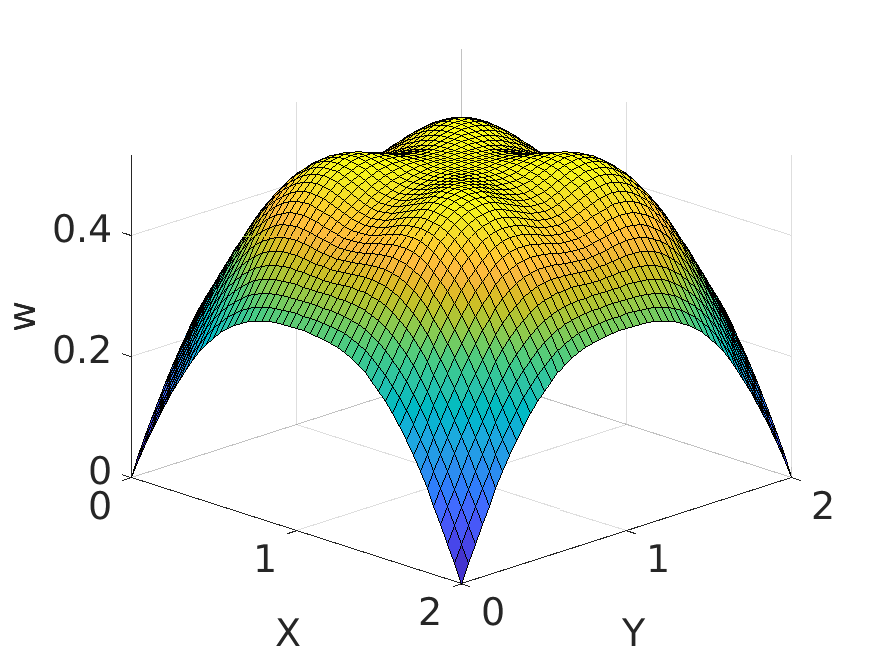}
\subcaption{$s=1,\Delta Q = \Delta P =-\frac{\pi}{3}$}
\end{subfigure}%
\hspace{1pt}
\begin{subfigure}{.24\linewidth}
  \centering
\includegraphics[scale=0.3]{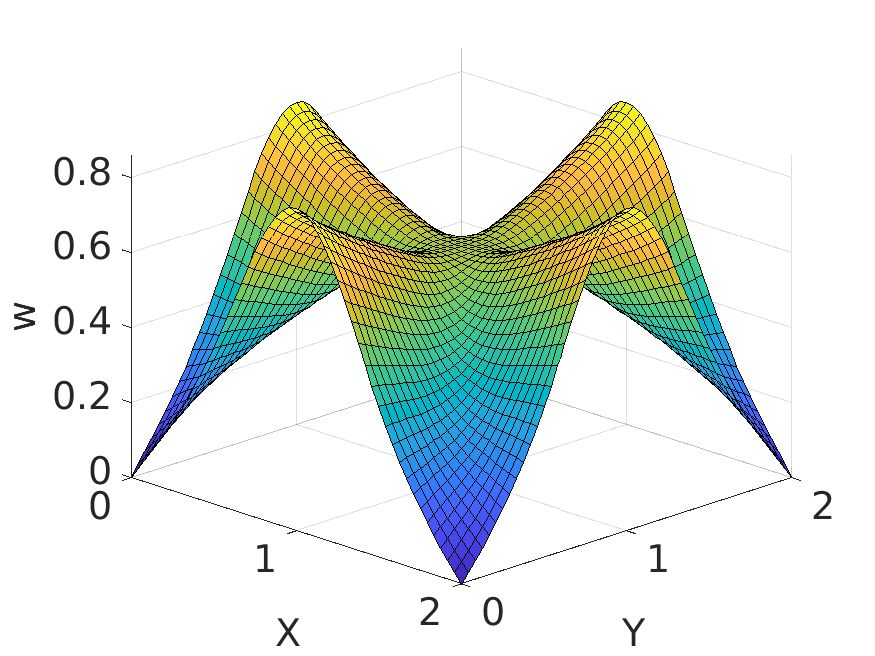}
\subcaption{$s=1,\Delta Q =\frac{\pi}{3}, \Delta P =-\frac{\pi}{3}$}
\end{subfigure}%
\hspace{1pt}
\begin{subfigure}{.24\linewidth}
  \centering
\includegraphics[scale=0.3]{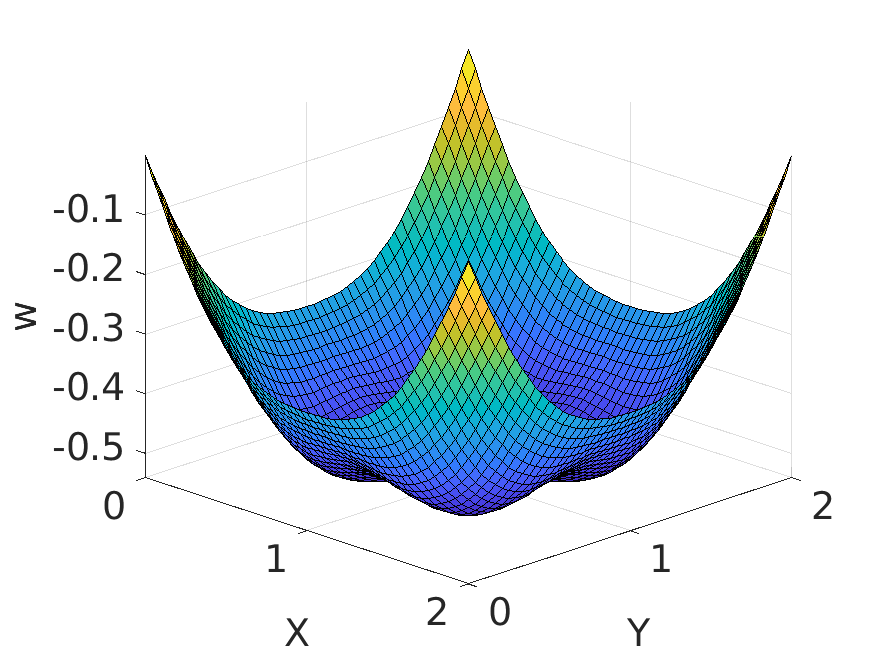}
\subcaption{$s=1,\Delta Q =-\frac{\pi}{3}, \Delta P =\frac{\pi}{3}$}
\end{subfigure}%

\begin{subfigure}{.24\linewidth}
  \centering
\includegraphics[scale=0.3]{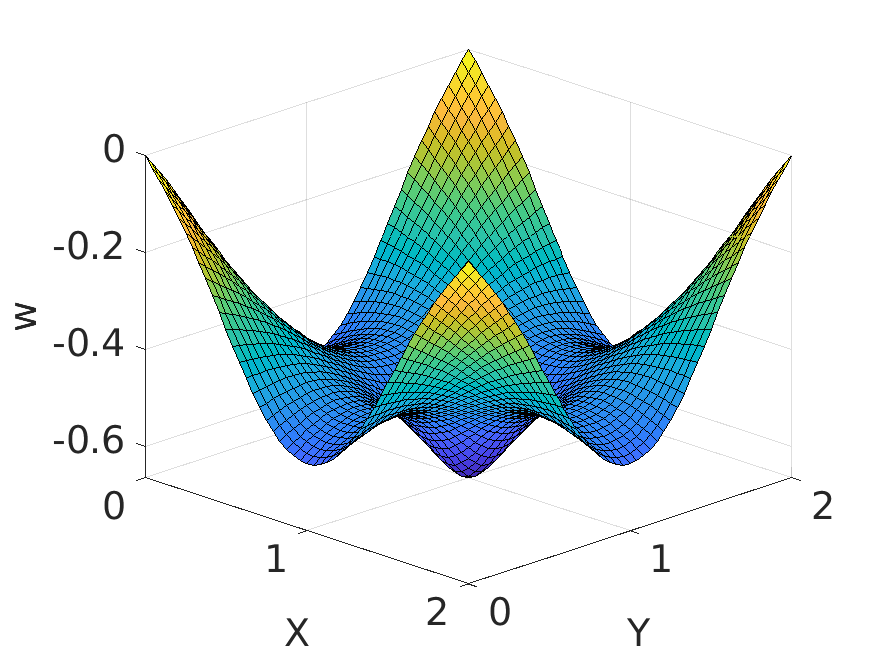}
\subcaption{$s=-1,\Delta Q =\frac{\pi}{3}, \Delta P =0$}
\end{subfigure}%
\hspace{1pt}
\begin{subfigure}{.24\linewidth}
  \centering
\includegraphics[scale=0.3]{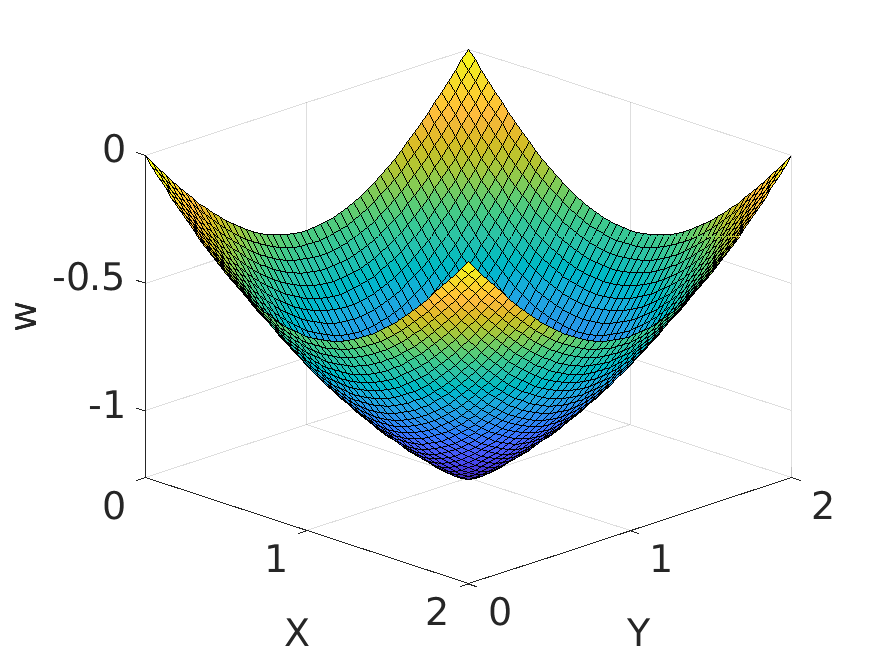}
\subcaption{$s=-1,\Delta Q =-\frac{\pi}{3}, \Delta P =0$.}
\end{subfigure}%
\hspace{1pt}
\begin{subfigure}{.24\linewidth}
  \centering
\includegraphics[scale=0.3]{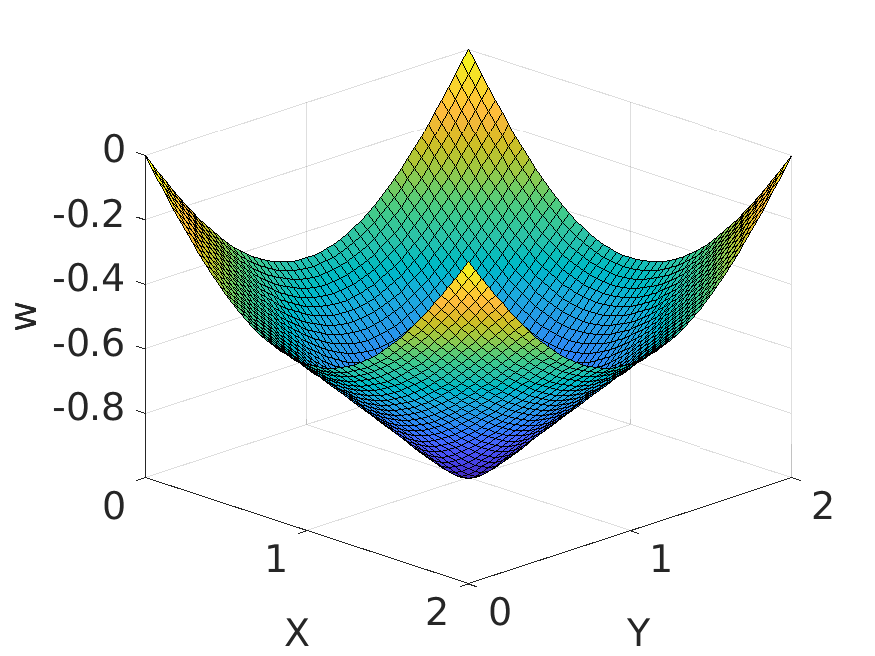}
\subcaption{$s=-1,\Delta Q =0, \Delta P =\frac{\pi}{3}$.}
\end{subfigure}%
\hspace{1pt}
\begin{subfigure}{.24\linewidth}
  \centering
\includegraphics[scale=0.3]{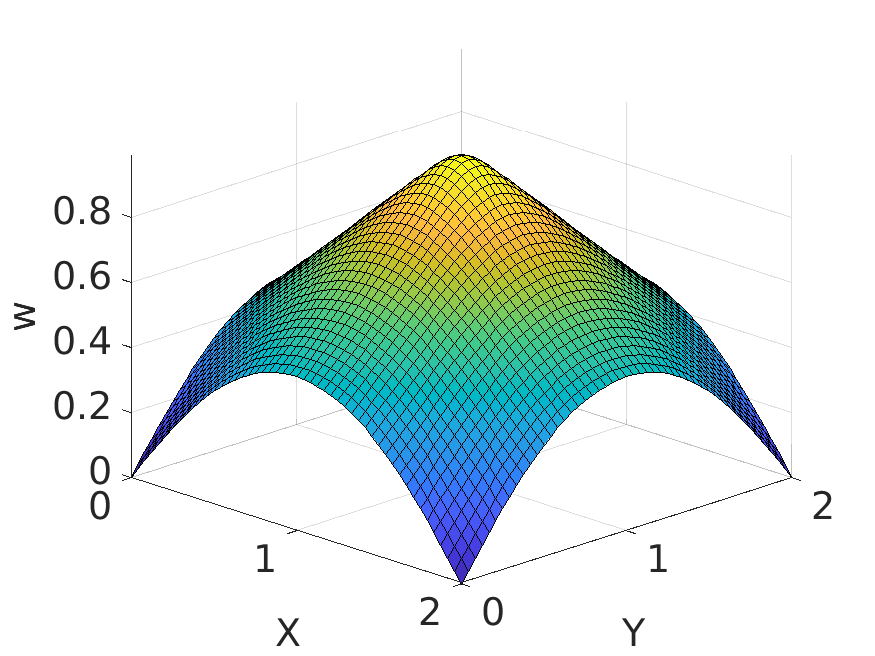}
\subcaption{$s=-1,\Delta Q =0, \Delta P =-\frac{\pi}{3}$}
\end{subfigure}%

\begin{subfigure}{.24\linewidth}
  \centering
\includegraphics[scale=0.3]{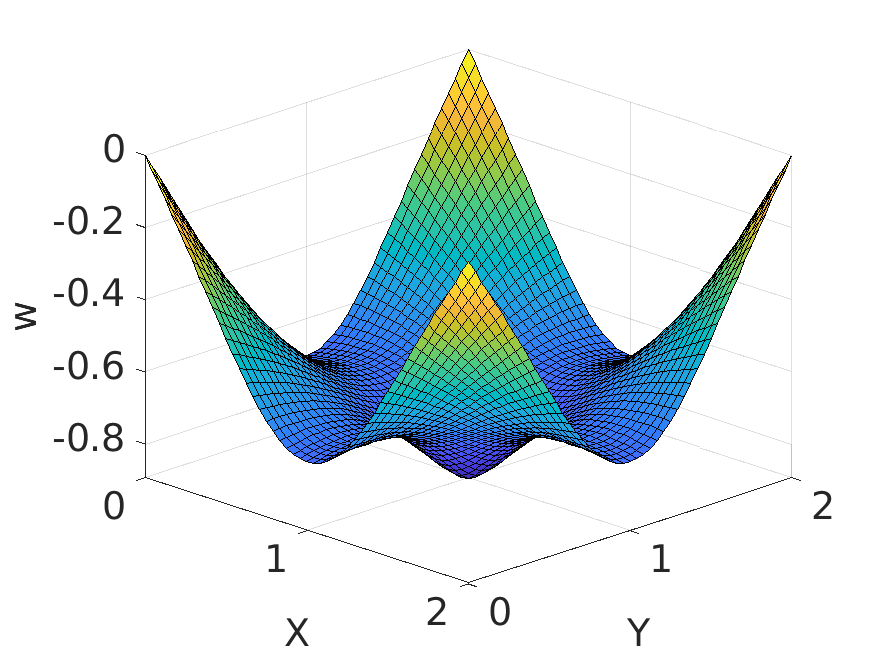}
\subcaption{$s=-1,\Delta Q = \Delta P =\frac{\pi}{3}$}
\end{subfigure}%
\hspace{1pt}
\begin{subfigure}{.24\linewidth}
  \centering
\includegraphics[scale=0.3]{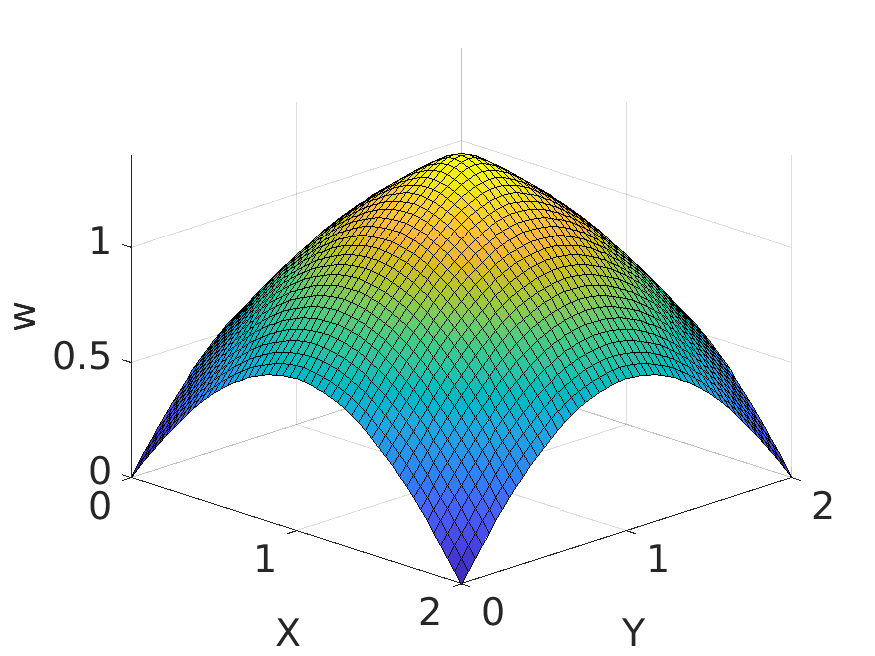}
\subcaption{$s=-1,\Delta Q =\Delta P =-\frac{\pi}{3}$}
\end{subfigure}%
\hspace{1pt}
\begin{subfigure}{.24\linewidth}
  \centering
\includegraphics[scale=0.3]{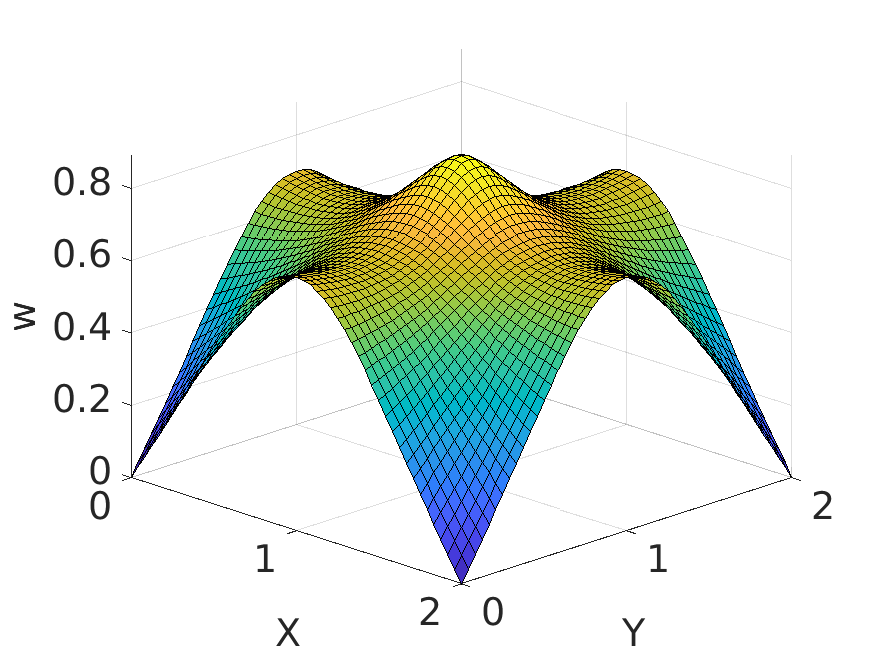}
\subcaption{$s=-1,\Delta Q =\frac{\pi}{3}, \Delta P =-\frac{\pi}{3}$}
\end{subfigure}%
\hspace{1pt}
\begin{subfigure}{.24\linewidth}
  \centering
\includegraphics[scale=0.3]{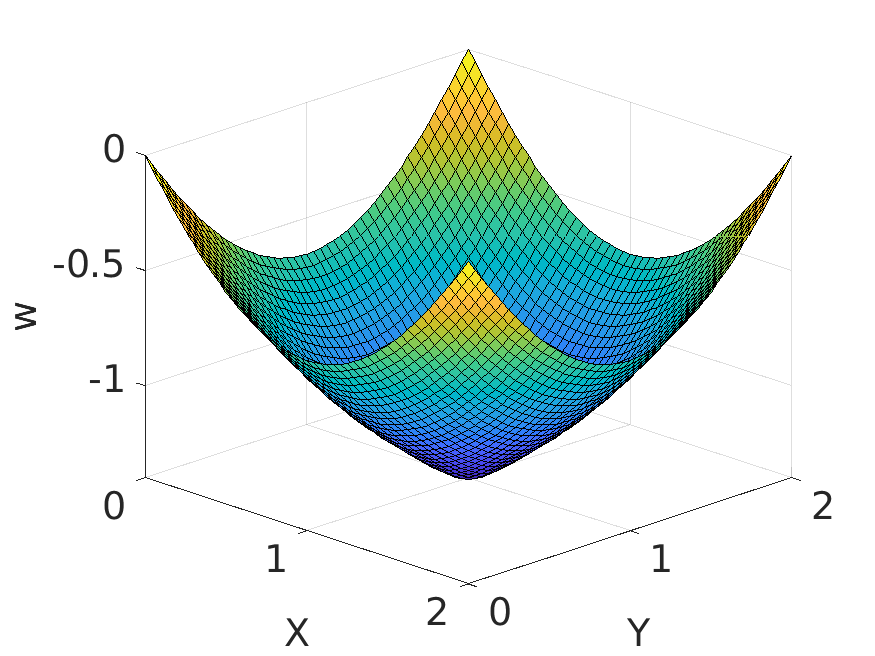}
\subcaption{$s=-1,\Delta Q =-\frac{\pi}{3}, \Delta P =\frac{\pi}{3}$}
\end{subfigure}%
\caption{The deformation $\text{w}$ for various coupled cases of isolated wedge disclinations and isotropic growth. The disclination density is of the form $\Theta=s\frac{\pi}{3}\delta_o$.}
\label{disclgrowth2}
\end{figure}

With the individual results in place, we study the coupling between disclination and growth. We consider several cases, all illustrated in Figure~\ref{disclgrowth2}, with various combinations of stretching and bending growth strains ($Q$ and $P$, respectively) coupled with either a negative wedge disclination ($\Theta = (\pi/3) \delta_o$), see Figure~\ref{disclgrowth2}(a-h), or a positive wedge disclination ($\Theta =- (\pi/3) \delta_o$), see Figure~\ref{disclgrowth2}(i-p), located at the centre of the square plate. In all the cases, the stress fields are distributed throughout the plate in addition to a concentration around the defect. The Gaussian curvature close to the centre of the plate is always dominated by the disclination whereas, away from the centre, it is influenced both by the defect and the growth strains. Also, as noted in the preceding paragraph, the plots with $\Delta P =0$ can equivalently appear with $\text{w}$ replaced by $-\text{w}$, i.e., after a reflection about the XY-plane. The presence of bending growth strain therefore fixes the direction (in terms of the orientation of the plate) into which the plate deforms.
The coupling between the disclination and the stretching growth strain $Q$, with bending growth strain absent, is captured in Figure~\ref{disclgrowth2}(a,b,i,j). Clearly, whenever $\Theta$ and $\Delta Q$ are of the same sign, as in Figure~\ref{disclgrowth2}(a,j), the deformation is enhanced (compared to the those in Figure~\ref{disclgrowth1}); the folds in Figure~\ref{disclgrowth2}(a) become more prominent and the cone in Figure~\ref{disclgrowth2}(j) more spherical. On the other hand, when the strength of the disclination and the magnitude of  $\Delta Q$ are of opposite sign, as in Figure~\ref{disclgrowth2}(b,i), the deformations are partially screened and there is a competition between saddle and spherical shape in Figure~\ref{disclgrowth2}(b) and between cone and a saddle-like shape in Figure~\ref{disclgrowth2}(c). Of course, the extent to which the changes in shapes will take place depends on the relative magnitude of defect and growth strain strengths. Nevertheless, the morphological richness that can be achieved by combining defect singularities and growth is evident from these simple simulations. On adding the effect of the bending growth strain $P$, the deformations are further enhanced in all the cases and, most importantly, the direction into which the plate will deform gets fixed, see Figure~\ref{disclgrowth2}(e-h,m-p). This is also evident when disclinations are coupled with bending growth strain, with stretching growth strain absent, as in Figure~\ref{disclgrowth2}(c,d,k,l).

\section{Conclusion} \label{conc}
A geometric theory of defects was used to derive inhomogeneous F{\"o}ppl-von K{\'a}rm{\'a}n equations for materially uniform, isotropic, and elastic shallow shells. The inhomogeneities were given in terms of defect densities arising from dislocations, disclinations, point defects, and metric anomalies. These equations provided the necessary governing equations for the micromechanical problem of determining deformation and stresses in a defective 2D surface in the absence of any external interactions of the surface. The general theory incorporated significantly wider class of defects than those present in the related literature and also coupled metric anomalies, such as those arising due to thermal or growth strains, with the defect distribution. The utility of the derived framework was illustrated through several numerical examples in the context of F{\"o}ppl-von K{\'a}rm{\'a}n plates. These included various configurations of an isolated edge dislocation (whose defect line and Burgers vector both lie within the plate surface) resulting into interesting folding patterns with negligible stress concentrations. Additionally, multiple simulation results were presented to highlight the coupling between a single wedge disclination, piercing through the plate, and a given distribution of isotropic stretching and bending growth strains. The present work provides a starting point to systematically study the mechanics of defects in curved surfaces, particularly for problems related to buckling, shape transformations, and defect interactions. Our work can also be extended to study the problem of defect dynamics which will involve kinetic relations for defect evolution and additional compatibility relations between spatial and temporal gradients of defect densities.

\section*{Acknowledgement} 
We gratefully acknowledge several insightful discussions with Mr. Animesh Pandey. AG acknowledges the financial support from SERB (DST) Grant No. CRG/2018/002873 titled ``Micromechanics of Defects in Thin Elastic Structures". 

\appendix

\section{The incompatible Gauss and Codazzi-Mainardi equations} \label{incompgcm}
The components $\tilde R_{ijkl}$ of the Riemann-Christoffel curvature tensor, associated with the Levi-Civita connection $S^k_{ij}$ such that $\tilde R_{klj}{}^i= S^i_{lj,k}- S^i_{kj,l}+S^h_{lj} S^i_{kh}-S^h_{kj} S^i_{lh}$ and $\tilde R_{klj}{}^i = g^{im} \tilde R_{kljm}$, and the components $\tilde \Omega_{kljp}$ of the material curvature are related to each other as~\cite[p.~141]{schouten}:
\begin{equation}
\tilde R_{ijkl}=\tilde \Omega_{ijkl} -2\tilde \nabla_{[i}\tilde W_{j]kl} -2\tilde W_{[i|m l|}\, \tilde W_{j]k}{}^m,
\label{identity1}
\end{equation}
where the tensor field  $\tilde W_{ij}{}^k(\theta^\alpha,\zeta)$  (with $\tilde W_{ijk} =\tilde W_{ij}{}^l g_{lk}$) is given additively as $\tilde W_{ij}{}^k =\tilde C_{ij}{}^k+ \tilde M_{ij}{}^k$, where $\tilde C_{ijk} =  \big(-\tilde T_{i k j} +  \tilde T_{kji} -  \tilde T_{ji k}\big)$ are components of the contorsion tensor and
$\tilde M_{ijk} = \frac{1}{2}\big( \tilde Q_{i k j} -  \tilde Q_{kji} +  \tilde Q_{ji k} \big)$ are components of the metric anomaly tensor~\cite{rcgupta17}. The respective surface restrictions are defined as $W_{ij}{}^k(\theta^\alpha)=\tilde W_{ij}{}^k(\theta^\alpha,\zeta=0)$, etc. Hence, 
\begin{equation}
C_{ijk} =  \big(-T_{i k j} +  T_{kji} -  T_{ji k}\big)~\text{and}~M_{ijk} = \frac{1}{2}\big( Q_{i k j} -  Q_{kji} +  Q_{ji k} \big). \label{contnonmet}
\end{equation}
From the inherent skew symmetry in $T_{ijk}$ and the symmetry in $Q_{ijk}$, it follows that $C_{i(jk)} = 0$ and $M_{[ij]k}=0$. As a result, $C_{ijj}=0$ (no summation on $j$).
The restriction of \eqref{identity1} to the mid-surface ($\zeta=0$) yields  \cite{rcgupta17}
\begin{subequations}
\begin{align}
& k_{1212}+ \text{det} [b_{\alpha \beta}]=I_1, \label{strain-incompatibility-relations1} \\
&-\nabla_2 b_{1 1}+\nabla_1 b_{1 2}=I_2, \label{strain-incompatibility-relations2}\\
&-\nabla_2 b_{12}+\nabla_1 b_{2 2}=I_3, ~\text{and} \label{strain-incompatibility-relations3}\\
&I_{\rho \sigma}=0, \label{strain-incompatibility-relations4}
\end{align}
\label{strain-incompatibility-relations}%
\end{subequations}
where $k_{1212}$ is the only independent component of the Riemannian curvature associated with the 2D metric $a_{\alpha\beta}$, such that $k_{1212} = \frac{1}{2} e^{\alpha \beta} e^{\mu \lambda} a_{\alpha \mu,\beta \lambda} - a^{\alpha \beta} (s_{12\alpha}s_{12\beta} - s_{11\alpha}s_{22\beta})$, and
\begin{subequations}
\begin{align}
& I_1=  a\Theta-2\nabla_{[1}W_{2]12}+ 2b_{1[1}W_{2]32} +2b_{[1|2|}W_{2]13}  -2W_{[1|i 2|}\, W_{2]1}{}^i, \\
&I_2= -a\Theta^2-2\nabla_{[1}W_{2]1 3} +2b_{1[1}W_{2]33} -2b^\rho_{[1}W_{2]1\rho}  -2W_{[1|i 3|}\, W_{2]1}{}^i,\\
&I_3= a\Theta^1-2\nabla_{[1}W_{2]2 3} +2b_{[1|2|}W_{2]33} -2b^\rho_{[1}W_{2]2\rho}  -2W_{[1|i 3|}\, W_{2]2}{}^i,~\text{and} \\
&I_{\rho \sigma} = a e_{\alpha \rho} e_{\beta \sigma} \Theta^{\alpha \beta} - \nabla_{\rho}W_{3 \sigma 3} + \tilde{W}_{\rho \sigma 3 , 3}\big|_{\zeta=0} - b^\delta_\rho W_{3\sigma \delta} + b^\delta_\sigma W_{\rho \delta 3} - 2 W_{[\rho |k 3|} W_{3] \sigma}{}^k.
\end{align}
\label{incompatibilities-defn}%
\end{subequations}
In writing the above expressions we have used the definitions $\nabla_\alpha W_{\beta \gamma 3} = W_{\beta \gamma 3 , \alpha} - s^\delta_{\alpha \beta} W_{\delta \gamma 3} - s^\delta_{\alpha \gamma} W_{\beta \delta 3}$ and  $\nabla_\alpha W_{3 \gamma 3} = W_{3 \gamma 3 , \alpha} - s^\delta_{\alpha \gamma} W_{3 \delta 3}$. The through-thickness derivative $\tilde{W}_{\rho \sigma 3,3} = \tilde{C}_{\rho \sigma 3 , 3} + \tilde{M}_{\rho \sigma 3 , 3}$, at $\zeta=0$, in the expression of $I_{\rho \sigma}$, is not derivable in terms of the surface defect and metric anomaly densities; it represents the out-of-plane variation of the densities. 

The conditions $I_1=0$, $I_2=0$, and $I_3=0$, reduce Equations \eqref{strain-incompatibility-relations1}-\eqref{strain-incompatibility-relations3} to the well-known Gauss and Codazzi-Mainardi equations, which provide necessary and sufficient conditions on sufficiently smooth first and second fundamental forms ($a_{\alpha \beta}$ symmetric positive definite and $b_{\alpha \beta}$ symmetric) for them to correspond to a smooth simple-connected surface embeddable in $\mathbb{R}^3$; the natural configuration is then realizable as a connected isometric embedding of the mid-surface $\omega$ in $\mathbb{R}^3$. Equations \eqref{strain-incompatibility-relations1}-\eqref{strain-incompatibility-relations3}, when at least one of $I_i$ is not zero, are the \textit{incompatible} Gauss and Codazzi-Mainardi equations with defect densities and metric anomalies as sources of incompatibility. Equation  \eqref{strain-incompatibility-relations4} provides identities for the interdependence of various defect types and metric anomalies. These identities are in addition to the various restrictions imposed on the defect and metric anomaly fields as discussed in the subsequent appendix section.

Under the order assumptions stated in Section~\ref{defectorder}, we collect the leading order expressions for the components of the contorsion and metric anomaly tensors (introduced in \eqref{contnonmet}):
\begin{subequations}
	\begin{align}
	& C_{3\mu}{}^3=C_{3\mu 3}=2e_{\mu\nu}\,\alpha^{\nu 3}=O(\epsilon^{\frac{1}{2}}),~C_{33\mu}=-C_{3\mu3},~C_{33}{}^\mu=\delta^{\sigma\mu}C_{33\sigma},\\
	& C_{\mu\nu}{}^3 =C_{\mu\nu 3} =e_{\mu\nu} {J} +(e_{\mu\sigma} \delta_{\rho\nu}+e_{\nu\sigma} \delta_{\rho\mu})\alpha^{\sigma\rho}=O(\epsilon^{\frac{1}{2}}), ~C_{\mu 3\nu}=-C_{\mu\nu 3},~C_{\mu 3}{}^\nu=\delta^{\rho\nu}C_{\mu 3 \rho}\\
	& C_{3\mu\nu}=-e_{\mu\nu}{J}+(e_{\mu\sigma} \delta_{\rho\nu}- e_{\nu\sigma} \delta_{\rho\mu})\alpha^{\sigma\rho}=O(\epsilon^{\frac{1}{2}}),~C_{3\mu}{}^\nu=\delta^{\rho\nu}C_{3\mu\rho},\\
	&C_{\alpha\beta\mu} = J^\sigma\big( \delta_{\sigma\beta} e_{\mu\alpha}  + \delta_{\sigma\alpha} e_{\mu\beta} + \delta_{\sigma\mu} e_{\alpha\beta}\big)=O(\epsilon), ~C_{\alpha\beta}{}^\mu = \delta^{\mu\nu} C_{\alpha\beta\nu},\\
&  M_{3\alpha}{}^{\beta} = M_{\alpha 3}{}^\beta =-\delta^{\beta\nu} q'_{\nu\alpha}=O(\epsilon^{\frac{1}{2}}),~M_{3\alpha\beta}=M_{\alpha 3\beta}=\frac{1}{2}Q_{3\alpha\beta}=-q'_{\alpha\beta}=O(\epsilon^{\frac{1}{2}})\\
	& M_{\alpha\beta}{}^3 = M_{\alpha\beta 3} =q'_{\alpha\beta}=O(\epsilon^{\frac{1}{2}}),\\
	&  M_{\alpha\beta\mu} =  (q^0_{\alpha\beta,\mu} - q^0_{\mu\beta,\alpha} - q^0_{\alpha\mu,\beta})=O(\epsilon),~\text{and}~M_{\alpha\beta}{}^\mu = \delta^{\mu\nu}M_{\alpha\beta\nu}.
	\end{align}
		\label{ContNonmetreduced}%
\end{subequations}
The components not mentioned above are either identically zero or order $O(\epsilon^{\frac{3}{2}})$. Furthermore, consequent to Equations~\eqref{bprel3-cons1}, we can obtain $\tilde{M}_{\rho \sigma 3 , 3}\big|_{\zeta=0} = 2 q''_{\rho \sigma}$. Consequently, the incompatibility measures in~\eqref{incompatibilities-defn} are simplified as 
\begin{subequations}
	\begin{align}
	 I_1=& ~ \Theta-2 e_{\alpha \beta} J^\alpha_{,\beta} + e^{\alpha\beta}e^{\mu\nu} q^0_{\alpha\mu,\beta\nu}  - (J)^2 + \text{det}[b_{\alpha \beta}] \nonumber\\
	& + \left(b_{12} + q'_{12} + \alpha^{22} - \alpha^{11}\right)^2  - \left(b_{22} + q'_{22} - 2\alpha^{12}\right)\left(b_{11} + q'_{11} + 2\alpha^{21}\right)+ O(\epsilon^{\frac{3}{2}}), \label{I1}\\
	I_2=& ~ -\Theta^2 + J_{,1} +  \left(q'_{11} + 2\alpha^{21} \right)_{,2} - \left(q'_{12} +\alpha^{22} -\alpha^{11} \right)_{,1}+ O(\epsilon^{\frac{3}{2}}),~\text{and}\\
	I_3=&~ \Theta^1 + J_{,2} +  \left(q'_{12} +\alpha^{22} -\alpha^{11} \right)_{,2} -  \left(q'_{22} - 2\alpha^{12} \right)_{,1} + O(\epsilon^{\frac{3}{2}}),
	\end{align}
	\label{incompatibilities1-fvk}%
\end{subequations}
where the leading order term in $I_1$ is $O(\epsilon)$ while the leading order terms in $I_2$ and $I_3$  are $O(\epsilon^{\frac{1}{2}})$.
The leading order terms in the expressions on the left hand sides of \eqref{strain-incompatibility-relations1}-\eqref{strain-incompatibility-relations3} can be obtained using $a_{\alpha \beta} = \delta_{\alpha \beta} + \text{w}^0_{,\alpha}  \text{w}^0_{,\beta} + 2 E^p_{\alpha\beta} +o(\epsilon)$ and $b_{\alpha \beta} =  \text{w}^0_{,\alpha\beta} + \Lambda^p_{\alpha\beta}+o(\epsilon^{\frac{1}{2}})$; these are $O(\epsilon)$ for \eqref{strain-incompatibility-relations1}, and $O(\epsilon^{\frac{1}{2}})$ for \eqref{strain-incompatibility-relations2} and  \eqref{strain-incompatibility-relations3}. Combining the left hand and the right hand side leading order terms in  \eqref{strain-incompatibility-relations1}-\eqref{strain-incompatibility-relations3}, we obtain the strain incompatibility equations~\eqref{s-i-r-fvk}.

\section{Identities relating defect densities}\label{conslaws}
The components of the material curvature tensor (with no metrical disclinations), the torsion tensor, and the non-metricity tensor, as introduced in Section \ref{defectdef}, are such that $\tilde{\Omega}_{(ij)kl}= 0$, $\tilde{\Omega}_{ij(kl)}= 0$, $\tilde{T}_{(ij)}{}^k=0$, and $\tilde{Q}_{k[ij]}=0$. In the following we assume $Q_{ij3} = 0$ to be in conformity with the Kirchhoff-Love hypothesis. The components also satisfy the following system of differential equations \cite[pp. 144-146]{schouten}:
\begin{subequations}
	\begin{align}
	& 2\tilde T_{[jk}{}^l{}_{;i]}={\tilde \Omega}_{[ijk]}{}^l+4\tilde T_{[ij}{}^m\,\tilde T_{k]m}{}^l,\label{diff:1}\\
	& \tilde \Omega_{[jk|l|}{}^m{}_{;i]}=2 \tilde T_{[ij}{}^q\tilde \Omega_{k]ql}{}^m,~\text{and}\label{diff:2}\\
	& \tilde Q_{[j|kl|;i]} + \tilde T_{ij}{}^m\,\tilde Q_{mkl}=0.\label{diff:3}
	\end{align}
	\label{bp-identities}%
\end{subequations}
In these equations, the anti-symmetrization with respect to three indices is defined as
\begin{equation}
A_{[nml]\cdots}{}^{\cdots}=\frac{1}{6}\left(A_{nml\cdots}{}^{\cdots}+A_{lnm\cdots}{}^{\cdots}+A_{mln\cdots}{}^{\cdots}-A_{lmn\cdots}{}^{\cdots}-A_{nlm\cdots}{}^{\cdots}-A_{mnl\cdots}{}^{\cdots}\right).
\end{equation}
Clearly, $A_{[\alpha\beta\mu]\cdots}{}^{\cdots} = 0$ and $A_{[nnl]\cdots}{}^{\cdots} = 0$ (no summation on $n$).
The relations \eqref{diff:1} and \eqref{diff:2} are both pairwise skew with respect to indices $i$, $j$, and $k$; none of these three indices can therefore take identical values. Without loss of generality, we can choose $i=1$, $j=2$, and $k=3$. Consequently, \eqref{diff:1} yield three independent relations which, when restricted  to the mid-surface, are
\begin{subequations}
	\begin{align}
\frac{1}{\sqrt{a}} \tilde T_{12}{}^\lambda{}_{;3}\big|_{\zeta=0} + \alpha^{\rho \lambda}{}_{;\rho}
&=\frac{1}{2} \varepsilon_{\rho\sigma} a^{\rho  \lambda} (\Theta^{\sigma}-\Theta^{\sigma 3})
+2 \varepsilon_{\rho\sigma}( \alpha^{\rho\sigma} J^{\lambda} + J^{\rho} \alpha^{\sigma \lambda} + \alpha^{\rho \lambda}\alpha^{\sigma 3})+  \frac{1}{2} \alpha^{\rho \lambda}Q_{\rho\mu}{}^\mu~\text{and} \label{disloconslaw1}\\
\frac{1}{\sqrt{a}} \tilde T_{12}{}^3{}_{;3}\big|_{\zeta=0} +  \alpha^{\rho 3}{}_{;\rho}
&=-\frac{1}{2}\varepsilon_{\rho\sigma}\Theta^{\rho\sigma} +2 \varepsilon_{\rho\sigma}(\alpha^{\rho\sigma} {J} + J^{\rho} \alpha^{\sigma 3})+\frac{1}{2} \alpha^{\rho 3} Q_{\rho\mu}{}^\mu. \label{disloconslaw2}
\end{align}
\label{firstbprel-nonlin}%
\end{subequations}
In writing the above relations we have used $\alpha^{ij}{}_{;k} = \alpha^{ij}{}_{,k} + L_{km}^i \alpha^{mj} + L_{km}^j \alpha^{im}$, with $L_{km}^i$ and $L_{km}^j$ evaluated at $\zeta=0$. In the absence of dislocations, these relations would imply that $\Theta^\rho=\Theta^{\rho 3}$ and $\Theta^{\rho\sigma}=\Theta^{\sigma\rho}$, i.e., $\Theta^{ij}$ is symmetric.
On the other hand, \eqref{diff:2} yields nine independent equations. One of these, corresponding to $l=m=3$, is trivial. Among the rest, three are differential equations which, on restricting to the mid-surface, provide us with balance equations for the disclination density tensor:
\begin{subequations}
	\begin{align}
	&\frac{1}{\sqrt{a}} \varepsilon^{\nu \alpha} \tilde \Omega_{12\alpha 3;3}\big|_{\zeta=0} + \Theta^{\rho\nu}{}_{;\rho} =
2 \varepsilon_{\rho \sigma} \big(\alpha^{\rho\sigma} \Theta^{\nu} 
-\alpha^{\rho 3} \Theta^{\sigma\nu} 
+ J^\rho  \Theta^{\sigma\nu} \big)
+\Theta^{\rho\nu} Q_{\rho\mu}{}^\mu~\text{and} \label{disliconslaw1}\\
& -\frac{1}{aa^{11}} \tilde \Omega_{122}{}^1{}_{;3}\big|_{\zeta=0} +  \Theta^{\rho 3}{}_{;\rho}  =
2\varepsilon_{\rho \sigma} \big(\alpha^{\rho \sigma} \Theta  
-\alpha^{\rho 3} \Theta^{\sigma 3} 
+ J^\rho   \Theta^{\sigma 3}\big)
+\frac{1}{2}\Theta^{\rho 3} Q_{\rho\mu}{}^\mu, \label{disliconslaw2}
	\end{align}
	\label{secondbprel-nonlin}%
\end{subequations}
where we have used $\Theta^{ij}{}_{;k} = \Theta^{ij}{}_{,k} + L_{km}^i \Theta^{mj} + L_{km}^j \Theta^{im}$, with $L_{km}^i$ and $L_{km}^j$ evaluated at $\zeta=0$. The remaining five relations are algebraic restrictions on the components of disclination density and non-metricity tensors, evaluated at $\zeta=0$,
\begin{subequations}
	\begin{align}
	& \Theta^{\alpha 3} Q_{\alpha 1}{}^2 = 0,~\Theta^{\alpha 3} Q_{\alpha 2}{}^1 = 0, \\
	& \Theta^{\alpha 3} Q_{\alpha 1}{}^1 = \Theta^{\alpha 3} Q_{\alpha 2}{}^2,~\text{and}\\
	& \Theta^{\alpha 1} Q_{\alpha \beta}{}^2 = \Theta^{\alpha 2} Q_{\alpha \beta}{}^1.
	\end{align}
	\label{thirdbprel-nonlin}%
\end{subequations}
The third relation \eqref{diff:3}, when rewritten as
\begin{equation}
\big(\tilde Q_{jkl,i}+L^m_{jk}\tilde Q_{iml}+L^m_{jl}\tilde Q_{imk}\big)_{[ji]}=0,
\label{bprel2}
\end{equation}
can be used to obtain a simplified representation for the non-metricity tensor \cite{rcgupta16, rcgupta17, rcgupta20}. A direct substitution shows that $\tilde Q_{kij}=-2 \tilde q_{ij;k}$ is a non-trivial solution of \eqref{bprel2},
where $\tilde q_{ij}=\tilde q_{ji}$ are arbitrary symmetric functions also known as quasi-plastic strain fields. Consequently, we have
\begin{equation}
Q_{kij}(\theta^\alpha)=-2 \tilde q_{ij;k}\big|_{\zeta=0}.
\label{irrot-1}
\end{equation}
The identities, analogous to Equations \eqref{firstbprel-nonlin}, \eqref{secondbprel-nonlin}, \eqref{thirdbprel-nonlin}, and \eqref{irrot-1},  for the full 3D framework are given in Equation (28) of \cite{rcgupta16}. 

\begin{figure}[t!]
	\centering
\includegraphics[scale=0.6]{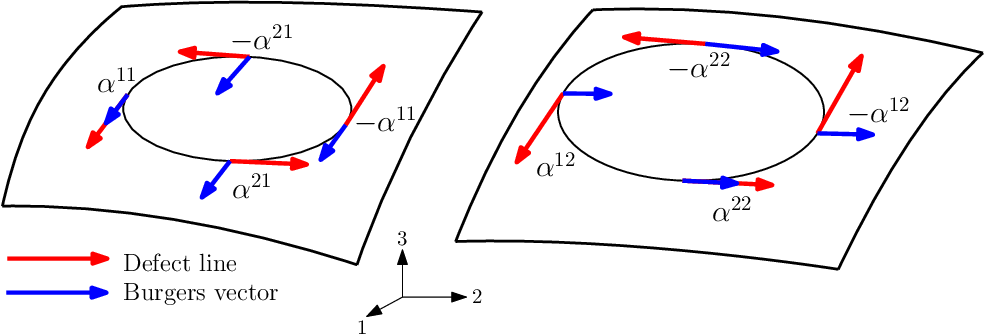}
	\caption{Conservation of $\alpha^{\rho 1}$ and $\alpha^{\rho 2}$.}
	\label{conservation}
\end{figure}

We now recover the leading order relations from the above identities. The identities \eqref{thirdbprel-nonlin} are order $O(\epsilon^{\frac{3}{2}})$ and will not be useful for the present discussion. Assuming the through thickness derivatives in relations \eqref{firstbprel-nonlin} and \eqref{secondbprel-nonlin} (the first term in all the equations) to be order $O(\epsilon)$, we obtain the leading order balance laws (of order $O(\epsilon^{\frac{1}{2}})$) as 
\begin{equation}
	\alpha^{\rho \lambda}{}_{,\rho}
	=\frac{1}{2} e_{\rho\sigma} \delta^{\rho  \lambda}(\Theta^{\sigma}-\Theta^{\sigma 3}),~
	\alpha^{\rho 3}{}_{,\rho}
	=-\frac{1}{2} e_{\rho\sigma}\Theta^{\rho\sigma},~\text{and}~\Theta^{\rho k}{}_{,\rho}
	=0.
	\label{dislconslaws}
\end{equation}
Therefore, skew part of the disclination density $\Theta^{ij}$ acts as a source/sink for the out-of-surface dislocation densities  $\alpha^{\mu k}$. If $\Theta^{ij}$ is symmetric then the defect lines corresponding to such dislocations must either leave the mid-surface or form loops on the mid-surface; see Figure~\ref{conservation} for an illustration. Conversely, if the out-of-surface dislocation fields are necessarily divergence free (conserved) then the disclination density is symmetric. The relation \eqref{dislconslaws}$_3$ similarly represents the conservation of  out-of-surface disclination density fields. The out-of-surface disclinations represented by  $\Theta^{\mu k}$ must either leave the mid-surface or form loops on the mid-surface. The disclination conservation law implies that there exists a vector potential field, with components $\vartheta^k$ of order $O(\epsilon^{\frac{1}{2}})$, such that $\Theta^{\mu k}= e^{\mu\nu} \vartheta^k{}_{,\nu}$.
Combining this with \eqref{dislconslaws}$_1$, we write 
\begin{equation}
\Theta^{2}=2\alpha^{\rho 1}{}_{,\rho}-  \vartheta^3{}_{,1}~\text{and}~ \Theta^{1}=-2 \alpha^{\rho 2}{}_{,\rho}+  \vartheta^3{}_{,2}. 
\label{bprel21-fvk22}
\end{equation}
If we assume that $\tilde{C}_{\rho \sigma 3 , 3}\big|_{\zeta=0} = O(\epsilon)$ then the leading order terms in Equation \eqref{strain-incompatibility-relations4}, which are $O(\epsilon^{\frac{1}{2}})$, yield $e_{\alpha \rho} e_{\beta \sigma} \Theta^{\alpha \beta} - 2 e_{\sigma \nu} \alpha^{\nu 3}{}_{,\rho} =0$ or equivalently
\begin{equation}
\Theta^{11} = -2\alpha^{13}{}_{,2},~\Theta^{22} = 2\alpha^{23}{}_{,1},~\Theta^{12} = -2\alpha^{23}{}_{,2},~\text{and}~\Theta^{21} = 2\alpha^{13}{}_{,1}. \label{disl1213disc12}
\end{equation}
The components $\vartheta^1$ and $\vartheta^2$ of the potential can therefore be identified with  $-2\alpha^{13}$ and $-2\alpha^{23}$, respectively, without any loss of generality.

\section{The variational principle and the numerical method}
\label{numerics}

The numerical examples in Sections \ref{disexp} and \ref{disclexp} consider a square plate domain $\omega$. In all the problems we assume the free boundary conditions such that, on every point of the boundary $\partial \omega$,
\begin{subequations}
\begin{align}
 &\Phi=0, ~\nabla\Phi\cdot\boldsymbol{n}=0,\\
 &\boldsymbol{m}\cdot\boldsymbol{n}\otimes\boldsymbol{n}=0,~\text{and}~
 \nabla(\boldsymbol{m}\cdot\boldsymbol{n}\otimes\boldsymbol{t})\cdot \boldsymbol{t} +(\nabla\cdot \boldsymbol{m})\cdot\boldsymbol{n}=0,
 \end{align}
 \label{bc}%
\end{subequations}
where $\boldsymbol{m}$ is the moment tensor, constitutively given as $\boldsymbol{m}=-D\left((1-\nu)\nabla^2 \text{w} +\nu \Delta \text{w} \boldsymbol{1}\right)$, $\boldsymbol{t}$ is the unit tangent to the boundary, and $\boldsymbol{n}$ is the in-plane unit normal to the boundary. Whereas the first two conditions enforce that there are no net in-plane forces applied at any point of the boundary, the latter two ensure that there is no moment (about $\boldsymbol{t}$) and no transverse shear force, respectively, being applied at any point of the boundary. The complete boundary value problem for dislocations in plates is given by \eqref{Firstvk-onlyd}, \eqref{Secvk-onlyd}, and \eqref{bc}, where the evaluation of $\psi$ is to be done separately by solving the Poisson's equation \eqref{laplscrew} subjected to Dirichlet boundary condition $\psi = 0$ on $\partial \omega$. The complete boundary value problem for disclinations in plates is given by \eqref{Firstvk-discl}, \eqref{Secvk-discl}, and \eqref{bc}, where the evaluation of $\psi$ is to be done separately by solving the Poisson's equation \eqref{lapldiscl} subjected to Dirichlet boundary condition $\psi = 0$ on $\partial \omega$. 

 We solve the boundary value problems using a finite element methodology. We have developed our own code (using MATLAB R2021a; codes are included as supplementary files) using a mixed variational principle, according to which the governing equations appear as the stationary conditions of the functional \cite[p. 165]{washizu}
\begin{equation}
\begin{aligned}
\Pi (\text{w},\Phi)=&\frac{D}{2}\int_{\omega}\left( (\Delta \text{w} )^{2}-2(1-\nu)\text{det}(\nabla^2 \text{w})\right) dA  
 -\frac{1}{2E}\int_{\omega}\left( (\Delta \Phi )^{2}-2(1+\nu) \text{det}(\nabla^2 \Phi)\right) dA \\ 
&+ \frac{1}{2}\int_{\omega}\left( (\nabla^2 \Phi(\nabla \text{w} \times \mathbf{e}_{3}))\cdot (\nabla \text{w} \times \mathbf{e}_{3})\right) dA -\int_{\omega}\lambda^{p} \Phi dA + D \int_{\omega} \Omega^{p} \text{w} dA,
\end{aligned} 
\end{equation}
where $\lambda^p$ and $\Omega^p$ are expressed in terms of defect densities. The square plate domain is discretized using non-conforming C$^1$-continuous rectangular elements (following Reddy~\cite[Ch. 6]{reddy}) and the weak form of the variational principle is used to obtain a system of nonlinear algebraic equations. The algebraic equations are solved using an arc-length method which is able to trace the nonlinear equilibrium path through the limit point (including snap-back and snap-through). We note that the equations are nonlinear and hence the solutions obtained are not unique. Different solution paths can be traced depending on the initial guess of the parameters involved in the numerical procedure. All the solutions are stationary points of the functional $\Pi$ but not all are necessarily stable. The stable (metastable) solution corresponds to a point of global (local) minima in the strain energy landscape.

\end{document}